\documentclass[sigconf]{acmart}

\AtBeginDocument{%
  \providecommand\BibTeX{{%
    \normalfont B\kern-0.5em{\scshape i\kern-0.25em b}\kern-0.8em\TeX}}}

 \AtBeginDocument{
 \providecommand\BibTeX{{%
  Bib\TeX}}}

\copyrightyear{2023}
\acmYear{2023}
\setcopyright{rightsretained}
\acmConference[CCS '23]{Proceedings of the 2023 ACM SIGSAC Conference on Computer and Communications Security}{November 26--30, 2023}{Copenhagen, Denmark}
\acmBooktitle{Proceedings of the 2023 ACM SIGSAC Conference on Computer and Communications Security (CCS '23), November 26--30, 2023, Copenhagen, Denmark}\acmDOI{10.1145/3576915.3616586}
\acmISBN{979-8-4007-0050-7/23/11}

\makeatletter
\gdef\@copyrightpermission{
  \begin{minipage}{0.3\columnwidth}
   \href{https://creativecommons.org/licenses/by/4.0/}{\includegraphics[width=0.90\textwidth]{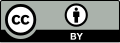}}
  \end{minipage}\hfill
  \begin{minipage}{0.7\columnwidth}
   \href{https://creativecommons.org/licenses/by/4.0/}{This work is licensed under a Creative Commons Attribution International 4.0 License.}
  \end{minipage}
  \vspace{5pt}
}
\makeatother

\usepackage{tikz}
\usetikzlibrary{plotmarks}
\usetikzlibrary{positioning, shapes.geometric}
\usepackage{amsmath,soul}
\usepackage[english]{babel}
\usepackage{graphicx,subfigure}
\usepackage{graphics}
\usepackage{amsmath}
\usepackage{url}
\usepackage{listings}
\usepackage{xcolor}
\usepackage[commandnameprefix=always, final]{changes}
\usepackage{xspace}
\usepackage{booktabs}
\usepackage{tablefootnote}
\usepackage{algorithm}
\usepackage{algorithmic}
\usepackage{adjustbox}
\usepackage{numprint}
\usepackage{tcolorbox}
\usepackage{multirow}
\usepackage{colortbl}
\usepackage{xintexpr, xinttools}

\newcommand{\para}[1]{\smallskip \noindent \textbf{#1}}
\newcommand{\parait}[1]{\smallskip \noindent \textit{#1}}

\newcommand\eg{\emph{e.g.},\xspace}
\newcommand\ie{\emph{i.e.},\xspace}
\newcommand\etc{\emph{etc}.\xspace}

\providecommand{\etal}{\emph{et al.}\xspace}

\expandafter\def\expandafter\UrlBreaks\expandafter{\UrlBreaks
  \do\a\do\b\do\c\do\d\do\e\do\f\do\g\do\h\do\i\do\j%
  \do\k\do\l\do\m\do\n\do\o\do\p\do\q\do\r\do\s\do\t%
  \do\u\do\v\do\w\do\x\do\y\do\z\do\A\do\B\do\C\do\D%
  \do\E\do\F\do\G\do\H\do\I\do\J\do\K\do\L\do\M\do\N%
  \do\O\do\P\do\Q\do\R\do\S\do\T\do\U\do\V\do\W\do\X%
  \do\Y\do\Z}

\begin{document}

\newcommand\name{\textsc{CookieGraph}\xspace}
\newcommand\webgraph{WebGraph\xspace}

\newcommand*\fullcirc[1][1ex]{\tikz\fill (0,0) circle (#1);} 
\newcommand{\halfcirc}{
\tikz
    {
    \node (s1) [circle, fill=white, minimum size=1ex] {};
    \node      [semicircle, fill=black, 
                inner sep=0pt, outer sep=0pt, 
                minimum size=1ex,
                at={(s1.center)}, 
                rotate=90] {};
     }
}

\newcommand{\ats}{ATS\xspace}
\newcommand{\atses}{ATSes\xspace}
\newcommand{\nonats}{Non-ATS\xspace}
\newcommand{\unknown}{Unknown\xspace}

\newcommand{\accuracy}{90.18}
\newcommand{\precision}{90.07}
\newcommand{\recall}{92.09}

\newcommand{\atscookie}{first-party \ats cookie\xspace}
\newcommand{\nonatscookie}{first-party \nonats cookie\xspace}
\newcommand{\atscookies}{first-party \ats cookies\xspace}
\newcommand{\nonatscookies}{first-party \nonats cookies\xspace}

\newcommand{\cblockcrawl}{\texttt{3P-Blocked}\xspace}
\newcommand{\callowcrawl}{\texttt{3P-Allowed}\xspace}

\definecolor{lightgray}{rgb}{0.95, 0.95, 0.95}
\definecolor{darkgray}{rgb}{0.4, 0.4, 0.4}
\definecolor{editorGray}{rgb}{0.95, 0.95, 0.95}
\definecolor{editorOcher}{rgb}{1, 0.5, 0} 
\definecolor{editorGreen}{rgb}{0, 0.5, 0} 
\definecolor{orange}{rgb}{1,0.45,0.13}		
\definecolor{olive}{rgb}{0.17,0.59,0.20}
\definecolor{brown}{rgb}{0.69,0.31,0.31}
\definecolor{purple}{rgb}{0.38,0.18,0.81}
\definecolor{lightblue}{rgb}{0.1,0.57,0.7}
\definecolor{lightred}{rgb}{1,0.4,0.5}

\def\bluecolor{\color{blue}}

\lstdefinelanguage{JavaScript}{
  morekeywords={typeof, new, true, false, catch, function, return, null, catch, switch, var, if, in, while, do, else, case, break},
  morecomment=[s]{/*}{*/},
  morecomment=[l]//,
  morestring=[b]",
  morestring=[b]'
}

\lstdefinelanguage{HTML5}{
  language=html,
  sensitive=true,	
  alsoletter={<>=-},	
  morecomment=[s]{<!-}{-->},
  tag=[s],
  otherkeywords={
  >,
	<!DOCTYPE,
  </html, <html, <head, <title, </title, <style, </style, <link, </head, <meta, />,
	</body, <body,
	</div, <div, </div>, 
	</p, <p, </p>,
	</script, <script,
  <canvas, /canvas>, <svg, <rect, <animateTransform, </rect>, </svg>, <video, <source, <iframe, </iframe>, </video>, <image, </image>, <header, </header, <article, </article
  },
  ndkeywords={
  =,
  charset=, src=, id=, width=, height=, style=, type=, rel=, href=,
  fill=, attributeName=, begin=, dur=, from=, to=, poster=, controls=, x=, y=, repeatCount=, xlink:href=,
  margin:, padding:, background-image:, border:, top:, left:, position:, width:, height:, margin-top:, margin-bottom:, font-size:, line-height:,
  transform:, -moz-transform:, -webkit-transform:,
  animation:, -webkit-animation:,
  transition:,  transition-duration:, transition-property:, transition-timing-function:,
  }
}

\lstdefinestyle{htmlcssjs} {%
  basicstyle={\footnotesize\ttfamily},   
  frame=b,
  identifierstyle=\color{black},
  keywordstyle=\color{blue}\bfseries,
  ndkeywordstyle=\color{editorGreen}\bfseries,
  stringstyle=\color{editorOcher}\ttfamily,
  commentstyle=\color{brown}\ttfamily,
  language=HTML5,
  alsolanguage=JavaScript,
  alsodigit={.:;},	
  tabsize=2,
  showtabs=false,
  showspaces=false,
  showstringspaces=false,
  extendedchars=true,
  breaklines=true,
  literate=%
  {Ö}{{\"O}}1
  {Ä}{{\"A}}1
  {Ü}{{\"U}}1
  {ß}{{\ss}}1
  {ü}{{\"u}}1
  {ä}{{\"a}}1
  {ö}{{\"o}}1
}


\newcommand{\redline}{\raisebox{2pt}{\tikz{\draw[-,red,solid,line width = 3pt](0,0) -- (5mm,0);}}}
\newcommand{\orangeline}{\raisebox{2pt}{\tikz{\draw[-,orange,solid,line width = 3pt](0,0) -- (5mm,0);}}}
\newcommand{\greenline}{\raisebox{2pt}{\tikz{\draw[-,green,solid,line width = 3pt](0,0) -- (5mm,0);}}}

\newcommand{\blackline}{\raisebox{2pt}{\tikz{\draw[-,black,solid,line width = 3pt](0,0) -- (5mm,0);}}}
\newcommand{\grayline}{\raisebox{2pt}{\tikz{\draw[-,gray,solid,line width = 3pt](0,0) -- (5mm,0);}}}
\newcommand{\lightgrayline}{\raisebox{2pt}{\tikz{\draw[-,lightgray,solid,line width = 3pt](0,0) -- (5mm,0);}}}

\makeatletter 
\newcommand{\linebreakand}{%
  \end{@IEEEauthorhalign}
  \hfill\mbox{}\par
  \mbox{}\hfill\begin{@IEEEauthorhalign}
}
\makeatother 

\title[\name: Understanding and Detecting First-Party Tracking Cookies]{\name: \\Understanding and Detecting First-Party Tracking Cookies}

\author{Shaoor Munir}
\email{smunir@ucdavis.edu}
\affiliation{%
	\institution{University of California, Davis}
	\country{Davis, CA, USA}
}
\author{Sandra Siby}
\email{s.siby@imperial.ac.uk}
\affiliation{%
	\institution{Imperial College London}
	\country{London, United Kingdom}
}

\author{Umar Iqbal}
\email{umar.iqbal@wustl.edu}
\affiliation{%
	\institution{Washington University in St. Louis}
	\country{St. Louis, MO, USA}
}
\author{Steven Englehardt}
\email{se@senglehardt.com}
\affiliation{%
	\institution{Independent Researcher}
	\country{Highland Park, NJ, USA}
}
\author{Zubair Shafiq}
\email{zubair@ucdavis.edu}
\affiliation{%
	\institution{University of California, Davis}
	\country{Davis, CA, USA}
}
\author{Carmela Troncoso}
\email{carmela.troncoso@epfl.ch}
\affiliation{%
	\institution{EPFL}
	\country{Lausanne, Switzerland}
}

\renewcommand{\shortauthors}{Shaoor Munir et al.}

\begin{abstract}
	As third-party cookie blocking is becoming the norm in mainstream web browsers, advertisers and trackers have started to use first-party cookies for tracking.
To understand this phenomenon, we conduct a differential measurement study with versus without third-party cookies.
We find that first-party cookies are used to store and exfiltrate identifiers to known trackers even when third-party cookies are blocked. 

As opposed to third-party cookie blocking, first-party cookie blocking is not practical because it would result in major breakage of website functionality. 
We propose \name, a machine learning-based approach that can accurately and robustly detect and block first-party tracking cookies.
\name detects first-party tracking cookies with \accuracy\% accuracy, outperforming the state-of-the-art CookieBlock by 17.31\%.
We show that \name is robust against cookie name manipulation, while CookieBlock's accuracy drops by 15.87\%.
While blocking all first-party cookies results in major breakage on 32\% of the sites with SSO logins, and CookieBlock reduces it to 10\%, we show that \name does not cause any major breakage on these sites.

Our deployment of \name shows that first-party tracking cookies are used on 89.86\% of the top-million websites.
We find that 96.61\% of these first-party tracking cookies are in fact ghostwritten by third-party scripts embedded in the first-party context. 
We also find evidence of first-party tracking cookies being set by fingerprinting scripts. 
The most prevalent first-party tracking cookies are set by major advertising entities such as Google, Facebook, and TikTok.
\end{abstract}

\ccsdesc[500]{Security and privacy~Privacy protections}
\ccsdesc[300]{Security and privacy~Usability in security and privacy}
\ccsdesc[500]{Security and privacy~Domain-specific security and privacy architectures}
\ccsdesc[300]{Computing methodologies~Classification and regression trees}

\keywords{cookies, machine learning, privacy, tracking, web security}

\maketitle

\section{Introduction}
\label{sec:introduction}
Major browser vendors such as Safari, Firefox, and Google Chrome have either blocked or are in the process of blocking \textit{third-party cookies} --- cookies set on domains that differ from the domain of the site visited by a user~\cite{JohnWebkit2020-full-cookie-blocking, firefox-total-cookie-blocking, edge-tracking-protection}.
Because third-party cookies are accessible across different sites that a user visits, they are used for cross-site tracking (i.e., linking a user's browsing activity across different sites).
Due to their ubiquitous use in tracking, the question arises as to how trackers will respond to third-party cookie blocking.
\textit{First-party cookies} --- cookies that are set on the same domain as that being visited by a user -- are of particular interest to advertisers and trackers because they will still be available in the face of third-party cookie blocking. 
However, since first-party cookies are only accessible from the setting domain, it remains to be seen how they can be used in lieu of third-party cookies for cross-site tracking.

Prior literature has shown that first-party cookies set by third-party scripts can be exfiltrated to tracking endpoints~\cite{Sanchez2021JourneyToCookies, ChenACM2021CookieSwapParty, Fouad20PixelsPETS}. 
Prior work has also shown that trackers use browser fingerprinting to re-spawn first-party cookies~\cite{Fouad22CookieRespawningPETS}.
Yet, there is no work studying the full spectrum of tracking possible through first-party cookies; and crucially, no countermeasures exist to specifically detect and block first-party tracking cookies.
To fill this gap, we first investigate the use of first-party cookies by known trackers and then use our findings to develop a machine-learning based approach, \name, to detect and block first-party tracking cookies.

We first perform a differential measurement study comparing the use of first- and third-party cookies on a 20\% sample of top-million websites across parallel crawls, with third-party cookies enabled and blocked. 
We show that third-party cookie blocking does not significantly impact the sharing of identifiers to known tracking endpoints because major trackers are already using first-party cookies.
Our analysis reveals that these trackers store identifiers in first-party cookies based on probabilistic and deterministic information.

Unlike third-party cookies, blocking all first-party cookies is not practical, as some of these cookies might be required for legitimate website functionality.
An alternative could be the use of privacy-enhancing request blocking tools \cite{Iqbal20AdgraphSP, Sjosten20UnderservedRegionsWWW, Siby22WebGraph} that would also block the cookies set by the requested resources. 
Unfortunately, our evaluation shows that these tools also cause breakage because tracking cookies are often set by domains that also set functional cookies.
Researchers have recently started to develop approaches to detect and block (both first-party and third-party) tracking cookies \cite{XuehuiACMWSC21CookieMonster,DinoUSENIX22CookieBlock}. 
However, these approaches rely on content-based features such as cookie names and values, which can lead to a high number of false positives (and consequently major website breakage) while also being susceptible to evasion~\cite{Siby22WebGraph}.

To address these limitations, we design and implement \name, a machine-learning approach to specifically detect first-party tracking cookies.
Instead of using content-based features, \name attempts to capture fundamental tracking behaviors exhibited by first-party cookies that we discover in our differential measurement study. 
\name is able to detect first-party tracking cookies with \accuracy\% accuracy, outperforming the state-of-the-art CookieBlock \cite{DinoUSENIX22CookieBlock} by 17.31\%. 
We also show that blocking all first-party cookies results in major breakage on 32\% of the sites with SSO logins, which is improved to 10\% by CookieBlock.
In contrast, \name does not cause any major breakage on these sites.
Moreover, \name is robust to evasion through cookie name manipulation, while CookieBlock's accuracy degrades by 15.87\%.

We deploy \name on a 20\% sample of the top-million websites to find 108,947 first-party tracking cookies on 89.86\% of the websites. 
The most prevalent first-party tracking cookies are set by major advertising entities, such as Google, Facebook, and TikTok, and then exfiltrated to a large number of other advertising and tracking endpoints.
We find that 96.61\% of the first-party tracking cookies are in fact ghostwritten by third-party scripts, 223 of which also conduct fingerprinting, that are served from a total of 2,099 distinct third-party domains.

In summary, our key contributions are as follows:
\begin{enumerate}
  \item We conduct a \textbf{large-scale differential measurement study} to understand the usage of first-party cookies by trackers when third-party cookies are blocked. Our analysis shows that {blocking third-party cookies does not reduce the number of tracking requests containing identifiers} and provides evidence that {trackers already use first-party cookies in lieu of third-party cookies for tracking.}
  
  \item We introduce \name, \textbf{a machine-learning based countermeasure} to detect and block first-party tracking cookies. \name captures fundamental tracking behaviors of first-party cookies that. \name {outperforms the state-of-the-art in terms of accuracy, robustness, and breakage minimization}. 
  
  \item We \textbf{deploy \name on a 20\% sample of the top-million websites} to measure the prevalence of first-party tracking cookies. We detect a total of 2,099 distinct domains that set first-party tracking cookies, including major advertising entities such as Google, and show that {fingerprinting scripts set first-party cookies on 1,908 sites.}
  
\end{enumerate}

\para{Paper Organization:}
The rest of this paper is organized as follows: Section \ref{sec:background} provides an overview of the recent developments and related work on cookies.
Section \ref{sec:threat_model} describes the threat model of first-party cookies.
Section \ref{sec:measurements} presents our differential measurement study to evaluate the impact of third-party cookie blocking on the use of first-party cookies by trackers.
Section \ref{sec:counter_measures} describes the design and evaluation of \name.
Section \ref{sec:deployment} presents results from our deployment of \name.
We discuss the limitations of \name in Section \ref{sec:limitations} and conclude in Section \ref{sec:conclusion}.

\section{Background \& Related Work}
\label{sec:background}
\subsection{Adoption of third-party cookies for tracking}
Cookies were originally designed to recognize returning users, \eg to maintain virtual shopping carts \cite{Montulli2013Cookies}. Soon, they were adopted by third-parties to track users across websites and serve targeted ads \cite{doubleclick}.
Early standardization efforts focused on limiting unintended cookie sharing across domains \cite{IETFRFC2109} and, despite well-known privacy concerns \cite{ThisCookieBug1995FinancialTimes}, largely ignored the misuse of cookies by third-parties for cross-site tracking.
Over the years, the use of third-party cookies for cross-site tracking has become prevalent \cite{FranziskaUSENIX12,AaronWWW2016EmpiricalStudyCookies,Sanchez2021JourneyToCookies, Dambra2022sallytrackers}.
Prior research shows that the vast majority of third-party cookies are set by advertising and tracking services (ATS) \cite{Dambra2022sallytrackers} and third-party cookies outnumber first-party cookies by a factor of two \cite{AaronWWW2016EmpiricalStudyCookies} -- and up to four when they contain identifiers \cite{Sanchez2021JourneyToCookies}.

\subsection{Countermeasures against third-party cookies}
\subsubsection{Safari} 
Since its inception in 2003, Safari has blocked third-party cookies from domains that have not been visited by the user as full-fledged websites \cite{safari-default-cookie}. 
In 2017, Safari introduced Intelligent Tracking Prevention (ITP).
ITP uses machine learning to automatically detect third-party trackers.
It revoked storage access from classified domains if users did not interact with them on a daily basis \cite{JohnWebkit2017ITP}.
Since 2017, ITP went through several iterations, i.e., ITP 1.1 \cite{JohnWebkit2017ITP-1-1}, ITP 2.0 \cite{Wilander2018ITP2}, ITP 2.1 \cite{JohnWebkit2019ITP-2-1}, ITP 2.2 \cite{JohnWebkit2019ITP-2-2} and ITP 2.3 \cite{JohnWebkit2019ITP-2-3}, eventually leading to full third-party cookie blocking \cite{JohnWebkit2020-full-cookie-blocking}.

\subsubsection{Firefox} Firefox experimented with third-party cookie blocking in 2013 \cite{Eich2013Cookie,Eich2013CookieClearinghouse}, but did not ship default-on third-party cookie blocking until the release of Enhanced Tracking Protection (ETP) in 2018 \cite{Nguyen2018ETP}.
ETP blocks third-party cookies based on a blocklist of trackers provided by Disconnect \cite{disconnect-tracking-list}.
As of 2022, Firefox has launched Total Cookie Protection (TPC) which partitions all third-party cookie access \cite{firefox-total-cookie-blocking}.
Partitioning ensures that cookies set by a third party on one site are distinct from those set by the same third-party on other websites, eliminating the third party's ability to track users across those websites.

\subsubsection{Internet Explorer and Microsoft Edge} 
Amongst the mainstream browsers that have deployed countermeasures against third-party cookies, Internet Explorer (IE) and Microsoft Edge have the most permissive protections.
IE blocked third-party cookies from domains that did not specify their cookie usage policy with the P3P response header \cite{ie6p3p}.
However, website owners often misrepresented their own cookie usage policies, which rendered P3P ineffective  \cite{leon2010token}.
Since 2019, Microsoft Edge has blocked access to cookies and storage in a third-party context from some trackers, based on Disconnect's tracking protection list \cite{edge-tracking-protection,edge-tracking-protection-doc,disconnect-tracking-list}.

\subsubsection{Chrome}
Google Chrome is the only mainstream browser that does not restrict third-party cookies in any way in its default mode.
In 2020, Google announced plans to phase out third-party cookies in Chrome by 2022 \cite{Schuh2020ChromeCookies}. 
However, the plan has been postponed several times and the latest timeline suggests the phasing out of cookies by late 2024 \cite{GooglePrivacySandbox}. 

\subsection{Adoption of first-party cookies for tracking}
While third-party cookies are widely considered as the main mechanism for cross-site tracking, trackers have also relied on first-party cookies for various forms of tracking, as described below.

\para{Same-site tracking.} As early as 2012, Roesner \etal \cite{FranziskaUSENIX12}, noted that third-party tracking scripts, embedded on the main webpage (\ie in first-party context), set first-party cookies.
First-party cookies enable \textit{same-site tracking}, where trackers can determine whether a user is revisiting a website or internal pages of a site.
While not as invasive as tracking users across different sites, significant information about a user can be gleaned from tracking their activity on the sites they frequent (\eg a social media or news site).

\para{Cross-domain same-site tracking.} First-party cookies can also be used for \textit{cross-domain same-site tracking}, where a website's cookies are shared by trackers to other domains. 
In 2020, Fouad \etal \cite{Fouad20PixelsPETS} found that trackers sync first-party cookies to several third-parties on as many as 67.96\% of the websites they tested.
In 2021, Chen \etal \cite{ChenACM2021CookieSwapParty} found that more than 90\% of the websites contain at least one first-party cookie that is set by a third-party script.
Similar to Fouad \etal, they found that at least one first-party cookie is exfiltrated to a third-party domain on more than half of the tested websites, raising concerns that these cookies might be used for tracking.
Sanchez \etal \cite{Sanchez2021JourneyToCookies} echoed these concerns, uncovering several instances where first-party cookies were ghostwritten by third-parties and then exfiltrated to other third-parties.
They conclude, through a large-scale measurement study of top websites and multiple case studies, that even after blocking third-party cookies, users are still at risk of first-party cookie based tracking.

Cross-domain sharing of first-party cookies presents a bigger privacy issue for users than same-site tracking.
While same-site tracking is only restricted to domains that are able to set first-party cookies, cross-domain sharing of first-party cookies allows other trackers, which are not collaborating with the first-party domains, to receive information about user activity.
This simplifies operations for trackers as instead of collaborating with each different publisher to set first-party cookies, they can instead leverage tracking cookies set by another tracker to monitor user activity.
With this practice, not only the third-party domains that are setting first-party cookies can track users' activities on the site, but tracking is also extended to other domains that receive these first-party cookies. 

\para{Cross-site tracking.} While third-party cookies have been used extensively in \textit{cross-site tracking}, \ie where a tracker links a user's activity across sites, the mechanisms by which first-party cookies are used in cross-site tracking have not been studied so far.
Oh \etal \cite{cartologychang2022ccs} investigated the sharing of first-party data with trackers in lieu of third-party cookie blockage, determining that identifiers such as email addresses were also shared to popular trackers.
Their experiments show that trackers make use of identifiers like email addresses to link user activity across different sites.
They make use of this knowledge to perform identity entanglement, where an attacker can make use of an email address or other identifiers to influence the advertisements shown to a victim.
This sharing of additional information when third-party cookies are blocked allows trackers to track users across different sites.

Previous research has also shown that it is non-trivial to generate first-party identifiers that are accessible across websites. 
Prior research has found that trackers often leverage browser fingerprinting to generate first-party tracking cookies \cite{Fouad22CookieRespawningPETS}.
Browser fingerprinting provides unique identifiers that are accessible across websites but drift over time \cite{laperdrix2016beauty}.
However, identifiers generated through browser fingerprinting can be stored in cookies that persist even after fingerprints change.
In addition to browser fingerprinting, several advertising and tracking services, such as Google Ad Manager \cite{google-ppid-about} and ID5 \cite{yieldbird-identity-guide}, specify in their documentation that they also use publisher-provided identifiers (PPIDs), such as email addresses, to set first-party cookies.

We note that techniques such as CNAME cloaking also allow advertisers and trackers to use first-party cookies.
However, as prior work has extensively studied first-party cookie leaks due to CNAME cloaking, we do not focus on CNAME cloaking in this paper.

\subsection{Countermeasures against first-party cookies}

\subsubsection{Deployed countermeasures}
Safari is the only mainstream browser that has deployed protections against first-party tracking cookies.
Safari's ITP expires first-party cookies and storage set by scripts in 7 days if users do not interact with the website \cite{safari-default-cookie}.
This limit is lowered to 24 hours if ITP detects link decoration being used for tracking \cite{safari-default-cookie}.
However, first-party cookie tracking does not require link decoration to be effective.
In cases where link decoration is not used, trackers can still track users within the 7-day window and beyond if users interact with the website within the 7-day window.

\subsubsection{Countermeasures proposed by prior research}
There exist two machine-learning based approaches to detect tracking cookies.
Hu et al.~\cite{XuehuiACMWSC21CookieMonster}'s approach uses sub-strings in cookie names (\eg track, GDPR) as features to detect first-party and third-party tracking cookies. 
Bollinger et al.~\cite{DinoUSENIX22CookieBlock} proposed CookieBlock. 
CookieBlock uses several cookie attributes such as the domain name of the setter, cookie name, path, value, expiration, \etc as features to detect first-party and third-party tracking cookies. 
These approaches rely on hard-coded content features, which makes them susceptible to adversarial evasions (as we show later in Section~\ref{sub:robustness_comparison}).
Moreover, these approaches mainly rely on self-disclosed cookie labels as ground truth, which can be unreliable \cite{Van2021Cookie}.

\subsubsection{Request blocking approaches}
Request blocking through browser extensions, such as Adblock Plus \cite{adblockplus-web}, and machine-learning-based tracker detection approaches proposed by prior research, \eg \cite{Siby22WebGraph}, can block first-party cookies set by tracking requests.
However, request blocking is prone to cause breakage because it blocks access to content or cookies that might be essential for website functionality. 
We confirm this is the case in Section \ref{sub:robustness_comparison})

\para{\textit{Unique focus of this paper.}}
Prior work has only incidentally measured the usage of first-party tracking cookies, and existing approaches to detect first-party tracking cookies are lacking.
In this paper, we fill this void by conducting a large-scale study to measure the prevalence of first-party tracking cookies and develop an accurate and robust machine-learning approach, \name, aimed at detecting first-party cookies.

\section{Threat Model}
\label{sec:threat_model}
In this section, we describe the threat model of tracking via first-party cookies.\footnote{This threat model is informed by prior literature \cite{Sanchez2021JourneyToCookies,ChenACM2021CookieSwapParty,cartologychang2022ccs,Fouad20PixelsPETS,Fouad22CookieRespawningPETS} and our case studies of popular tracking services described in Appendix \ref{app:case_studies}}

\parait{First- vs third-party cookies.}
Before describing the threat model, we define what we mean by first- and third-party cookies.
Cookies can either be set by the Set-Cookie HTTP response header or by using document.cookie in JavaScript.
Cookies set via response header from the \textit{same} domain as the first-party are \textit{first-party} cookies.
Similarly, cookies set via response header from a \textit{different} domain than the first-party are \textit{third-party} cookies.
When cookies are set by a script, their classification depends on whether the script is embedded in a first- or third-party execution context.
The cookies set by third-party scripts \textit{running in the first-party context} are \textit{first-party} cookies. 
The cookies set by third-party scripts \textit{running in a third-party context} (e.g., third-party iframes) are \textit{third-party} cookies. 
There are three main entities in this threat model: users (the victim), trackers (the adversary), and publishers.

\para{We assume that the user: }
\begin{itemize}
    \item visits different websites using one or more desktop/mobile devices that have distinct \textit{fingerprints} \cite{laperdrix2020browser}
    \item is not averse to logging in to those websites and providing PII (personally identifiable information) such as email addresses
    \item has third-party cookies disabled and first-party cookies enabled
\end{itemize}

\para{We assume that the publisher:}
\begin{itemize}
    \item controls the content on the site being visited by the user
    \item embeds the tracker in the first-party context, allowing the tracker to set first-party cookies 
    \item shares email and other deterministic identifiers (e.g., username, phone number) with the tracker, if provided by the user
\end{itemize}

\para{We assume that the tracker: }
\begin{itemize}
    \item is present in a first-party context on the publisher's site
    \item can set and read first-party cookies using document.cookie
    \item can collect information such as IP addresses, screen resolution \etc, which can be used to construct a device fingerprint
\end{itemize}

\begin{figure}[!htpb]
	\centering
	\includegraphics[width=1\linewidth]{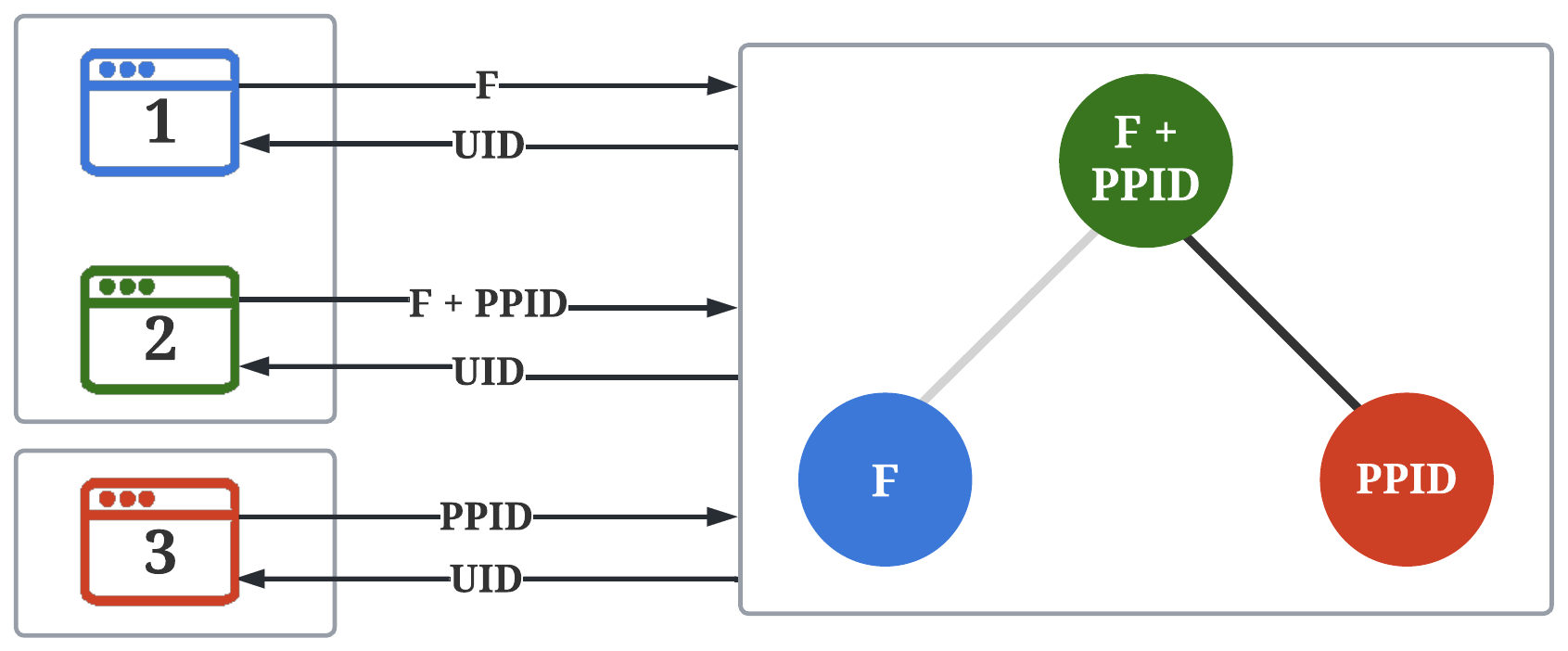}
	\caption{Cross-site tracking. The flow of information and identifiers through an identity graph for cross-site tracking. Initially, the user visits publishers 1 and 2 from one device. Tracker A, on publishers 1 and 2, collects and sends fingerprint $F$ to the identity graph. The identity graph returns a $UID$ for all the publisher visits, by matching fingerprints sent by each respective publisher. A publisher-provided ID, $PPID$, is also sent when visiting publisher 2. The user visits publisher 3 on a different device, thus tracker A is unable to construct a fingerprint that matches $F$. Publisher 3 sends a publisher-provided ID that matches $PPID$ provided by publisher 2. As a result, the identity graph matches and returns the same $UID$ for publisher 3. This ID is stored in first-party cookies on the user's device for each respective publisher.}
 \label{fig:idgraph}
\end{figure}

Trackers can use the information shared by the publisher, and the fingerprints collected by their own scripts to perform same-site, cross-domain same-site, and cross-site tracking, described below:
%

\para{Same-site tracking.} A user visits the same publisher's site multiple times.
During the first visit of the user, tracker A sets a first-party cookie on the user's device.
Upon subsequent visits by the user, tracker A can read the first-party cookie set and know that it is the same user who is revisiting the site.
When performing same-site tracking, tracker A is able to gather information about the user across the pages maintained by the same publisher.

\para{Same-site cross-domain tracking.} 
After setting a first-party cookie on a user's device, tracker A also shares the first-party cookie with a different tracker B that is not present in the first-party context (and thus is unable to set a first-party cookie of its own).
On each subsequent visit of the user, tracker A shares the first-party cookie and the pages visited by the user with tracker B.
Thus, without setting its own first-party cookie and directly colluding with the publisher, tracker B is also able to track the user's activity on the same site.

\para{Cross-site tracking.}
Consider a scenario in which a user visits three different sites (publishers 1, 2, 3) where tracker A is embedded in the first-party context.
The user visits sites 1 and 2 on one device and site 3 on a different device.
Publishers 2 and 3 ask the user for a deterministic identifier (\eg email address) which we denote as $PPID$ (Publisher-Provided ID).
Tracker A also constructs fingerprints on sites 1 and 2, denoted by $F_i$, where $i$ denotes the publisher visited.

When the user visits sites 1 and 2, tracker A collects fingerprints $F_1$ and $F_2$, which are the same (\ie $F_1 = F_2 = F$) as they are all constructed for the same device.
This allows tracker A to infer that the same user/device is visiting both sites. 
Tracker A links the deterministic identifiers and fingerprints belonging to the same user/device by constructing an \textit{identity graph} (refer to Appendix \ref{app:case_studies} for examples). 
The gray edge in Figure~\ref{fig:idgraph} shows the link in the identity graph constructed by tracker A for the fingerprints on sites 1 and 2.

The user then visits site 3 from a different device where tracker A is not able to construct the same fingerprint $F$. 
Publisher 3 asks the user for a deterministic identifier (\eg email address), which is the same as the $PPID$ provided by the user to publisher 2. 
Based on this additional information, tracker A can add a black edge to the identity graph.

Tracker A is finally able to connect all nodes in the identity graph to the user. 
Tracker A then assigns all connected nodes in the identity graph the same ID $UID$, which it can store in a first-party cookie on each of the sites.
On each subsequent visit by the user to any of the sites, tracker A can now simply read the first-party cookie containing $UID$.
Because $UID$ is the same across sites 1, 2, and 3, this allows tracker A to track the user across different sites.

\section{Measurements}
\label{sec:measurements}
In this section, we conduct a preliminary measurement study to investigate the usage of first-party cookies by advertising and tracking services (\ats) when third-party cookies are blocked.

\subsection{Data Collection}
\label{subsec:methodology}
\para{Data collection.}
We use OpenWPM (v0.17.0) and Firefox (v102) \cite{Englehardt16OpenWPMCCS} to crawl a sample of 20K out of the top-million websites.
To ensure that our crawls cover websites of variable popularity, we crawl the top 1K sites -- ensuring coverage of the most popular websites-- uniformly sample 9K sites from the sites ranked 1K-100K, and another 10K from sites ranked 100k-1M in the Tranco list \cite{pochat2018tranco}.
To capture behaviors that may be different in the landing and internal pages of a website \cite{aqeel2020landing}, we perform an interactive crawl that covers both kinds of pages.
Specifically, for each site, we crawl its landing page and then select up to 20 internal pages to visit at random.
We conduct four parallel crawls: two with third-party cookies enabled (\callowcrawl) and two with third-party cookies blocked (\cblockcrawl).
Parallelizing the crawls minimizes temporal variations across crawls and mitigates the effect of the dynamic behavior of websites.
We also turn off additional protections against tracking provided by Firefox
\cite{firefox-etp}.
We repeat failed crawls up to four times.
We successfully conducted the four parallel crawls for 99.31\% of the 20K websites.

\para{Labeling tracking activity.} 
To label tracking, we use EasyList \cite{easylist} and EasyPrivacy \cite{easyprivacy}.
Specifically, we use them to label requests as tracking (\ats) or not tracking (\nonats). 
We label a request as tracking (\ats) if its URL matches the rules in either one of the lists. 
Otherwise, we label it as not tracking (\nonats).

Since the basic premise of tracking is to identify users, we are particularly interested in sharing of identifiers in these tracking requests.
In line with prior work \cite{Iqbal22USENIXKhaleesi,Englehardt15CookiesICWWW}, we define identifiers as a string that is longer than 8 characters and matches the regex $[a-zA-Z0-9\_=-]$.
Using this definition, we look for identifiers in  URL query parameters \cite{RandallTrackers2022Arxiv} and cookie values \cite{Sanchez2021JourneyToCookies,Diaz2015CrossDeviceCookies,ChenACM2021CookieSwapParty,AaronWWW2016EmpiricalStudyCookies}.

\subsection{Tracking When Third-Party Cookies Are Blocked}
\label{subsec:tracking-before-after-blockage}
We first study whether blocking third-party cookies effectively eliminates \ats requests.
We compare the number of requests containing identifiers with and without third-party cookies.
\begin{figure}[!htpb]
	\centering
	\includegraphics[width=\linewidth]{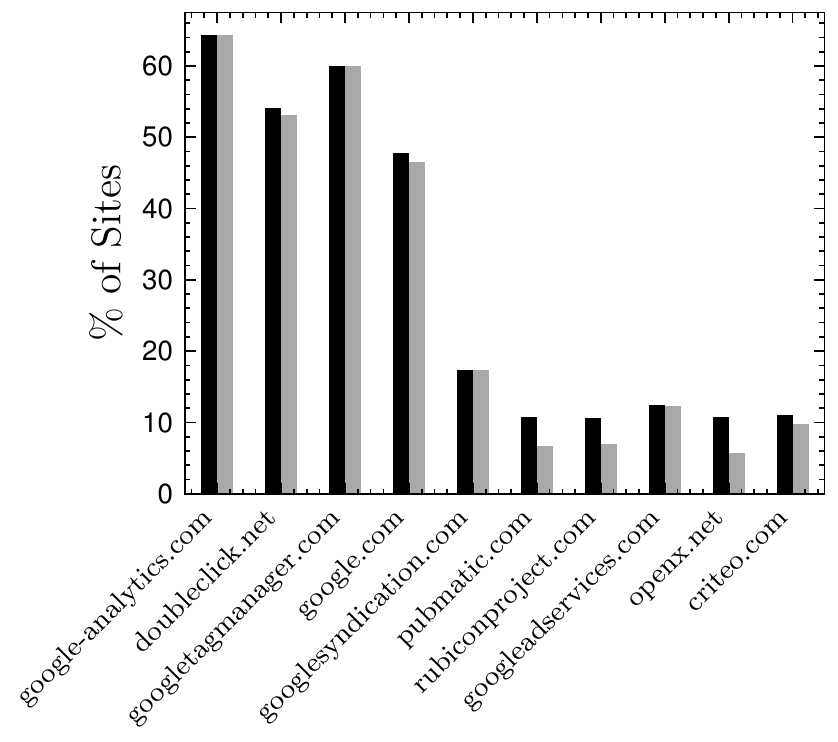}
	\caption[]{Presence of top-10 tracking domains. The plot shows the percentage of sites where at least one request containing an identifier is sent to a tracking domain.
		\\(\blackline) \callowcrawl: Third-party cookies allowed
		\\(\grayline) \cblockcrawl: Third-party cookies blocked}
	\label{fig:requests-breakdown-top-20}
\end{figure}

Table \ref{tab:request_stats} shows the average number of requests for two parallel crawls conducted with third-party cookies allowed and blocked.
We see that there is only a modest reduction in the overall number of \ats requests when third-party cookies are blocked.
The difference in the number of \ats requests containing identifiers is 10.82\%.
This is surprising because cookie syncing, which is widely used for same-site-cross-domain and cross-site tracking \cite{Fouad20PixelsPETS,Papadopoulos19cookiesyncing}, entails sharing third-party identifier cookies in query parameters \cite{Diaz2015CrossDeviceCookies,ChenACM2021CookieSwapParty,AaronWWW2016EmpiricalStudyCookies}.
With third-party cookies blocked, cookie syncing between third-parties cannot occur, and we would expect to see a larger drop in identifiers shared in \ats requests.
\textit{\textbf{We conclude that third-party cookie blocking does not effectively limit the exfiltration of identifiers to trackers.}}

\begin{table}[!htpb]
    \centering
    \caption{Average number of requests per site in \callowcrawl and \cblockcrawl configurations}
    \begin{tabular}{lccc}
    \toprule
    \textbf{Request Count} & \textbf{\callowcrawl} & \textbf{\cblockcrawl} &  \textbf{Change}\\
    \midrule
    Total &         771.47 &         766.43 &                 -0.65\% \\
    Tracking &          303.46 &         288.08 &                 -5.07\% \\
    Non-Tracking &         468.01 &         478.35 &                2.21\% \\
    Tracking with ID   &         126.43 &         112.75 &                 -10.82\% \\
    Tracking without ID  &         177.02 &         175.32 &                 -0.96\% \\
    \bottomrule
    \end{tabular}
    \label{tab:request_stats}
    \vspace{-10pt}
\end{table}
\begin{table}[!htbp]
    \centering
    \caption{Average number of first-party cookies per site in \callowcrawl and \cblockcrawl configurations}
    \resizebox*{\columnwidth}{!}{\begin{tabular}{lccc}
    \toprule
        \textbf{1P Cookie Count} & \textbf{\callowcrawl} & \textbf{\cblockcrawl} & \textbf{Change} \\
        \midrule
        Total                      &         132.65 &          137.73 &              -3.84\% \\
        Set by Trackers                &         109.74 &         114.53 &              -4.36\% \\
        Set by Non-Trackers            &          22.90 &          23.20 &              -1.31\% \\\
        Set by Trackers with ID   &           64.09 &          66.37 &             -3.55\% \\
        Set by Trackers without ID &          45.64 &          48.15 &              -5.50\% \\
        \bottomrule
    \end{tabular}}
    \label{tab:cookie_stats}
    \vspace{-10pt}
\end{table}
\begin{figure}[!htpb]
	\centering
	\includegraphics[width=\linewidth]{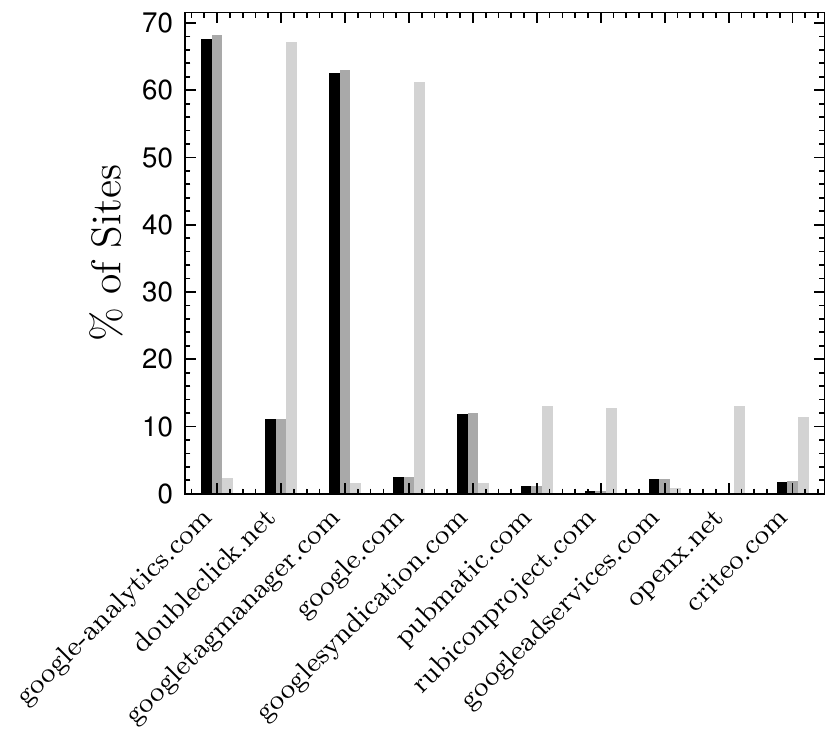}
	\caption[]{Comparison of percentage of sites on which first-party and third-party identifier cookies are set by \ats domains.
		\\(\blackline) first-party identifier cookies set when third-party cookies are allowed
		\\(\grayline) first-party identifier cookies set when third-party cookies are blocked
		\\(\lightgrayline) third-party identifier cookies set when third-party cookies are allowed}
	\label{fig:cookie-count-domain}
\end{figure}

Next, we analyze whether third-party cookie blocking disparately impacts different \ats domains (eTLD+1).
Figure \ref{fig:requests-breakdown-top-20} plots the percentage of sites with at least one \ats request with identifiers.
Six of the top-10 \ats domains, all owned by Google, show only a negligible reduction in the number of \ats requests with identifiers when  third-party cookies are blocked.
In contrast, three other \ats domains, owned by Pubmatic, Rubicon, and OpenX, show a significant reduction.
\vspace{-10pt}

\subsection{Tracking Through First-Party Cookies}
\label{subsec:tracking-analysis-first-party-cookies}
Table \ref{tab:request_stats} shows that even after blocking third-party cookies, there is only a small decrease in ATS requests containing identifiers (10.82\%).
The identifiers in these \ats requests are likely originating from some storage mechanism other than third-party cookies.
Since recent prior work has shown that \ats are increasingly using first-party cookies \cite{Sanchez2021JourneyToCookies, ChenACM2021CookieSwapParty}, we next investigate whether first-party cookies are being used in lieu of third-party cookies to circumvent third-party cookie blockage.

We first compare the average number of first-party cookies in \callowcrawl and \cblockcrawl crawls in Table \ref{tab:cookie_stats}.
We observe only a minor difference in the average number of first-party cookies set with third-party cookies allowed/blocked.
It is also noteworthy that 83.15\% of the first-party cookies are set by \ats scripts.
A further 57.94\% of them are identifier cookies.
We conclude that the vast majority of first-party cookies are in fact set by \ats and that they are not significantly impacted by third-party cookie blocking.

Next, we compare the setting of first- and third-party identifier cookies by \ats domains (eTLD+1 of the setting script URL) to understand if first-party cookie usage is equally prevalent across different \atses. 
Figure \ref{fig:cookie-count-domain} plots the percentage of sites where at least one first-party and/or third-party identifier cookie is set by a top-10 \ats domain.

For the six Google-owned \ats domains, which showed a negligible difference in requests containing identifiers after blocking third-party cookies, there is also little to no change in the use of first-party identifier cookies across both crawls.
\chdeleted{These domains do not set a large number of third-party identifier cookies, even when those are allowed, which likely explains why they were not impacted by third-party cookie blocking.}
\chadded{Two of these domains (\textit{doubleclick.net} and \textit{google.com}) set a large number of third-party cookies, while the remaining four (\textit{google-analytics.com}, \textit{googletagmanager.com}, \textit{googlesyndication.com}, and \textit{googleadservices.com}) do not.
We show in section \ref{sec:deployment} that the former two domains are a popular syncing endpoint for first-party cookies set by the other \ats domains, which explains their resilience to third-party cookie blocking.
%
}

On the contrary, the other set of \ats domains for which we observe a reduction of identifiers (i.e., Pubmatic, Rubicon, and OpenX) do use more third-party identifier cookies than first-party identifier cookies when third-party cookies are authorized.
This observation also explains the drastic drop in the number of requests containing identifiers to these other \ats domains after blocking third-party cookies in Figure \ref{fig:requests-breakdown-top-20}.
\textit{\textbf{We conclude that trackers that are not affected by third-party cookie blocking are using first-party cookies as a replacement.}} 

\begin{figure*}[!htbp]
	\centering
	\includegraphics[width=\linewidth]{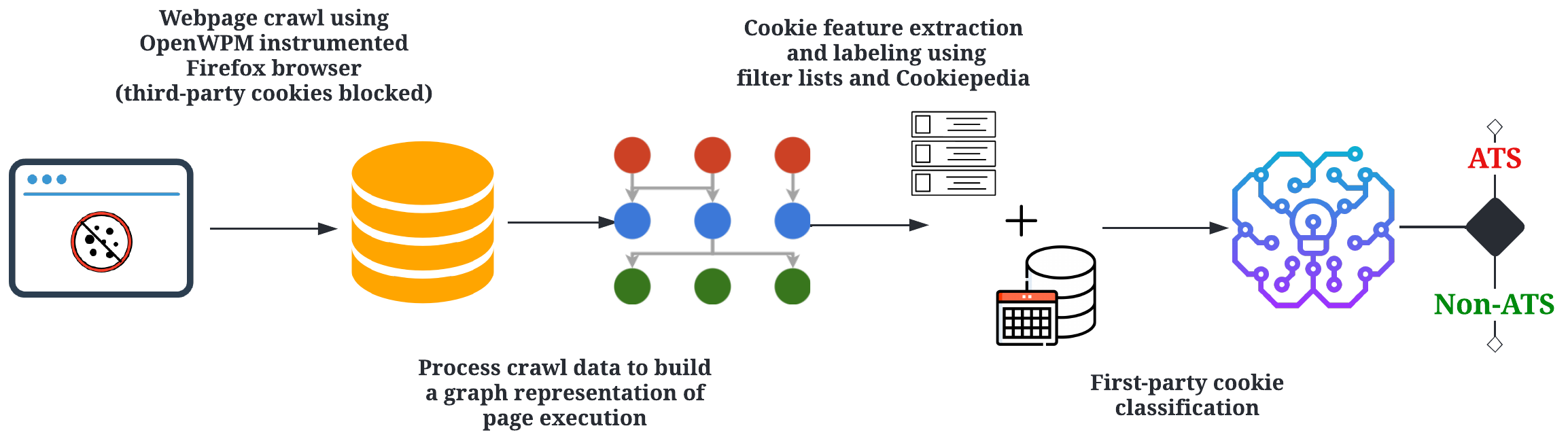}
	\caption{Overview of \name pipeline: (1) Webpage crawl using an instrumented browser; (2) Construction of a graph representation to represent the instrumented webpage execution information; (3) Feature extraction for graph nodes that represent first-party cookies; and (4) Classifier training to detect \atscookies.}
	\label{fig:classification_overview}
    \vspace{-10pt}
\end{figure*}

\subsection{Takeaway}
Our differential measurement study reveals that third-party cookie blocking does not effectively prevent tracking. 
There is only a negligible reduction in the exfiltration of identifiers to trackers when third-party cookies are blocked.
We find that this is because \atses use first-party cookies in lieu of third-party cookies.

We also find that the impact of third-party cookie blocking is not uniform across different trackers.
Some \ats domains show more reduction in the exfiltration of identifiers than others.
This disparity exists because some trackers only use first-party cookies regardless of the availability of third-party cookies; while others are using both first-party and third-party cookies to store identifiers.

\section{\name: Detecting First-Party Tracking Cookies}

\label{sec:counter_measures}
In this section, we describe \name, a graph-based machine learning approach to detect \atscookies.
\name creates a graph representation of a webpage's execution based on HTML, network, JavaScript, and storage information collected by an instrumented browser.
In this graph, first-party cookies are represented as storage nodes.
\name extracts distinguishing features of these cookies and uses a random forest classifier to detect \atscookies.
Figure \ref{fig:classification_overview} provides an overview of \name's pipeline.

\subsection{Design and Implementation}
\label{subsec:graph}
\para{Browser instrumentation.} \name uses our extended version of OpenWPM \cite{Englehardt16OpenWPMCCS} to capture webpage execution information across HTML, network, JavaScript, and the storage layers of a webpage.
Our analysis reveals significant usage of localStorage in addition to cookies.
In total, we found 217,444 unique first-party cookie names and 99,682 unique localStorage names.
In addition to this, we found 13,571 instances where the same first-party cookie was also stored in local storage. Thus, we use the term ``storage'' to refer to both cookies and localStorage.
In most cases, the description for cookies is also applicable to localStorage and vice versa.

\name captures HTML elements created by scripts, network requests sent by HTML elements (as they are parsed) and scripts, responses received by the browser, exfiltration/infiltration of identifiers in network requests/responses, and read/write operations on the browser's storage mechanisms.
\begin{figure*}[!htpb]
  \centering
    \includegraphics[width=\linewidth]{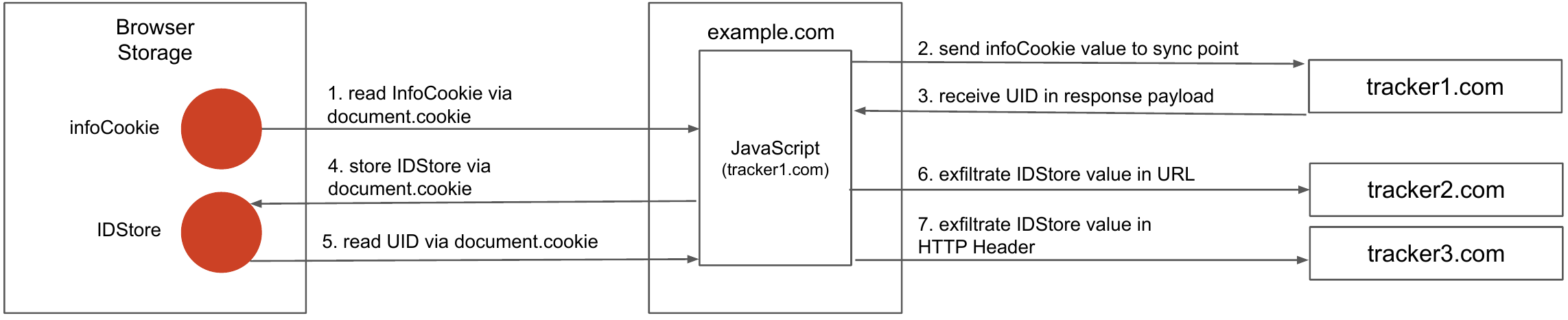}
  \caption{Example scenario to illustrate \name's graph construction (shown in Figure~\ref{fig:cookiegraph_example}).}
  \label{fig:cookiegraph_example}
  \vspace{-10pt}
\end{figure*}

\begin{figure}[!htpb]
  \centering
    \includegraphics[width=\linewidth]{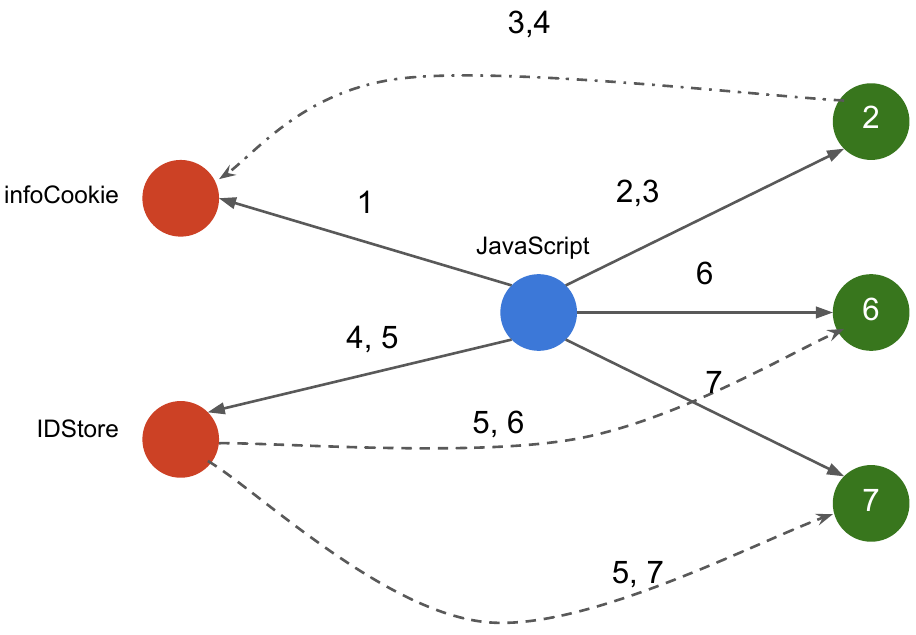}
  \caption{Graph representation of Figure~\ref{fig:cookiegraph_example} in \name. \protect\tikz{\protect\draw[fill=black!40!green] circle(1.0ex);} network nodes, \protect\tikz{\protect\draw[fill=black!40!blue] circle(1.0ex);} script nodes, and \protect\tikz{\protect\draw[fill=black!40!red] circle(1.0ex);} storage nodes. While the solid lines show the interactions of the script nodes with the storage and request nodes, the dashed (- - -) and dotted (. . .) lines represent the exfiltration and infiltration edges that are captured by \name.}
  \label{fig:cookiegraph_representation}
  \vspace{-10pt}
\end{figure}

\para{Graph construction.}
The nodes in \name's graph represent HTML elements, network requests, scripts, and storage elements.
When localStorage and first-party cookie nodes share the exact same name, \name considers them as one storage node.
\name's edges represent a wide range of interactions among different types of nodes \eg scripts sending HTTP requests, scripts setting cookies \etc
In addition to interactions considered by prior work \cite{Siby22WebGraph}, \name incorporates edges that model the actions associated to tracking using first-party cookies.
We identify these actions from the result of our measurement study in Section~\ref{sec:measurements}, and the case studies described in Appendix \ref{app:case_studies}.
Cookies are typically set with the values \textit{infiltrated} with HTTP responses and are \textit{exfiltrated} via URL parameters and request headers or bodies; \name captures infiltrations and exfiltrations by linking the script-read/write cookies in the first-party execution context to the requests of reader/writer script that contains those cookie values.
In addition to plain text cookie values, \name also monitors Base64-, MD5-, SHA-1-, and SHA-256- encoded cookie values in URLs, headers, request, and response bodies.
\name tracks the value of each cookie and associates the relevant interaction (exfiltration or infiltration) to the element that initiated the interaction.
Because of our focus on identifiers, \name only captures cookie values that are at least 8 characters long (but it would be trivial to extend it to consider smaller cookie values).
Figure~\ref{fig:cookiegraph_example} illustrates how \name creates a graph representation.
In this example, a third-party script from \textit{tracker1.com} executes in a first-party context on the webpage, \textit{example.com}. 
The script first reads \texttt{infoCookie} (1), which contains tracking information such as the publisher ID and a user signature. 
Then, the script sends the content of the cookie to \textit{tracker1.com}'s sync endpoint via an HTTP POST request (2).
The endpoint returns a user ID (UID) in the response body (3), which is stored in both a first-party cookie and localStorage named \texttt{IDStore} (4).
At a later point, the script reads the value from \texttt{IDStore} (5) and exfiltrates the UID to two other tracking endpoints: to \textit{tracker2.com} via a URL parameter (6) and to \textit{tracker3.com} via an HTTP header (7).

Figure~\ref{fig:cookiegraph_representation} shows the graph representation that \name generates for the execution of the example script.
The nodes in the graph represent the script, the storage, and the network endpoints.
The edge numbers show the actions performed in Figure~\ref{fig:cookiegraph_example}.
The dotted and dashed lines in the graph show the infiltration and exfiltration behaviors captured by \name.
\name is not only able to capture the interactions of the script with the storage and the network endpoints, but is also able to precisely \emph{link exfiltration and infiltration of the first-party cookie} via an edge from the cookie node to the endpoint.
%

\para{Feature extraction.}
We use \name's representation to extract two kinds of features.

\textit{Structural} features represent relationships between nodes in the graph, such as ancestry information and connectivity.
Structural features capture the relationships between the first-party cookie nodes and scripts on the page.
For example, how many scripts interacted with a cookie or whether a script that interacted with a cookie also interacted with other cookies.

\textit{Flow} features represent \atscookie behavior.
We extract three types of flow features.
First, we count the number of times a cookie was read or written.
Second, we count the number of times a cookie was infiltrated via HTTP responses or exfiltrated via URL parameters, request headers, or request bodies.
Third, features related to the setter of the cookie.
Concretely, whether the setter's domain also acted as an end-point for other cookie exfiltrations, and whether the setter's domain was involved in redirect chains (since redirects are commonly used in tracking).
The intuition behind the third category of features is that domains involved in setting \atscookies are also involved in sharing information with other \atses.

\name does \emph{not} use content features, such as cookie names, as they can be trivially modified to evade detection \cite{Siby22WebGraph, Iqbal22USENIXKhaleesi}.
\vspace{-10pt}

\subsection{Evaluation}
\label{subsec:classification}

Similar to previous work on graph-based webpage modelling~\cite{Iqbal20AdgraphSP, Siby22WebGraph}, we use a random forest classifier to distinguish between \ats and \nonats cookies.
We first train and test the accuracy of this classifier on a carefully labeled dataset.
Then, we deploy it on our 20K website dataset.

\subsubsection{Ground truth labeling}
\label{subsub: ground truth}
We use two complementary approaches to construct our ground truth for first-party \ats cookies. 
We represent each first-party cookie as a cookie-domain pair since the same cookie name can occur on multiple sites.

\para{Filter lists.}
We rely on filter lists \cite{easylist, easyprivacy} as previous work has found them to be reasonably reliable in detecting \ats endpoints \cite{Iqbal20AdgraphSP, Siby22WebGraph}.
Filter lists are designed to label resource URLs, rather than cookies. 
We adapt them to label cookies by assigning the label of a particular resource to all the cookies set by that resource.
Since both \ats and \nonats cookies can be set by the same resource, this labeling procedure could result in a non-trivial number of false positives.
To limit the number of false positives in our ground truth, we only label \nonats cookies based on filter lists: i.e., if a script that sets a cookie is not marked by \textit{any} of the filter lists, we label these cookies as \nonats.
Conservatively, if any one of the filter lists marks the cookie's setter as \ats, we label the cookie as \unknown. 

\para{Cookiepedia.}
Inspired by prior work \cite{DinoUSENIX22CookieBlock}, we use Cookiepedia~\cite{cookiepedia} as an additional source of cookie labels.
Cookiepedia is a database of cookies maintained by a well-known Consent Management Platform (CMP) called OneTrust \cite{hils2020measuringcmp,DinoUSENIX22CookieBlock}.
For each cookie/domain pair, Cookiepedia provides its purpose, defined primarily through the cookie integration with OneTrust.
Each cookie is assigned one of four labels: strictly necessary, functional, analytics, and advertising/tracking.
As Cookiepedia-reported purposes are self-declared, we adopt a conservative approach: we only label a cookie-domain pair as \ats if a cookie's purpose is declared as advertising/tracking or analytics in a particular domain. 
If the declared purpose is strictly necessary or functional, we label the cookie as \unknown, as the cookie might have been, mistakenly or intentionally, mislabeled. 

We combine the results of the labeling approaches to obtain a final label for the cookies. 
If both approaches label a cookie as \unknown, its final label is \unknown. 
If only one of the approaches has a known label, this is the final label.
If Cookiepedia marks a cookie as \ats and filter lists mark it as \nonats, we give precedence to the Cookiepedia label and assign the final label as \ats because websites are unlikely to self-declare their \nonats cookies as \ats.

Using this labeling process, 82,098 out of 304,162 (26.99\%) first-party cookie and domain pairs have a known (\ats or \nonats) label and the rest are labeled as \unknown.
We observe that cookies set by the same script across two different sites are often labeled \ats in one instance and \unknown in another instance because Cookiepedia does not have data for the latter.
As it is unlikely that an \ats script changes purpose across sites, we propagate the \ats label to all instances set by the same script. 
Using this label propagation, we label 37.92\% of the data, with 53,183 (46.10\%) \ats and 62,184 (53.90\%) \nonats labels.

\subsubsection{Classification} 
\label{subsec:classification-evaluation}

We train and test the classifier on the labeled dataset using standard 10-fold cross-validation.
We ensure that there is no overlap in the websites used for training and testing in each fold.
Similar to Section \ref{subsec:graph}, we limit the classifier to cookies whose value is at least 8 characters long. 
The classifier has \precision\% precision and \recall\% recall, with an overall accuracy of \accuracy\%, indicating that the classifier is successful in detecting \ats cookies.

\subsection{Feature Analysis}
\label{app:feature_analysis}
\begin{figure}[!htpb]
  \centering
  \includegraphics{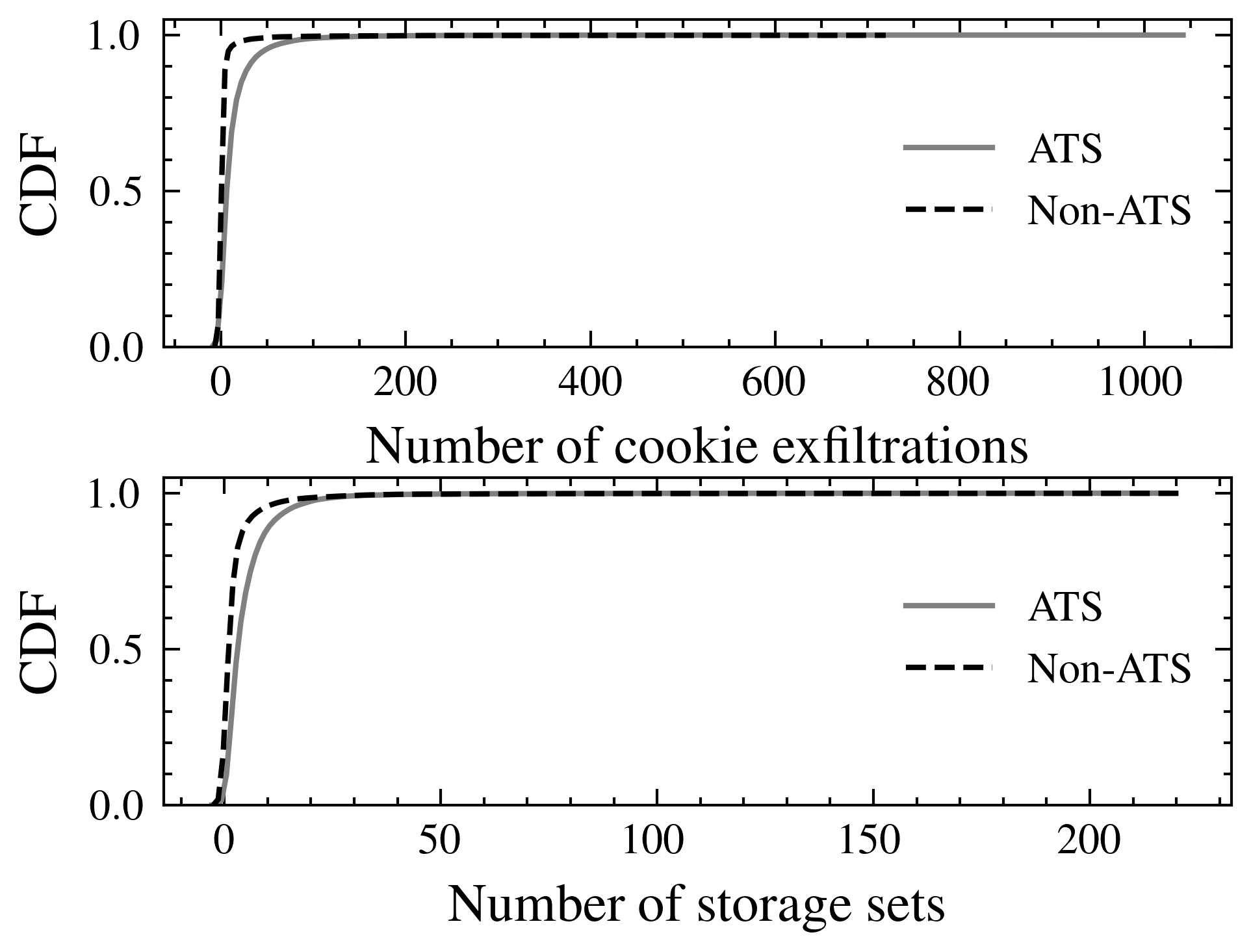}
  \caption{Feature distribution of cookie exfiltrations (top) and storage sets (bottom) for \ats and \nonats cookies. \ats cookies are exfiltrated and set more than \nonats cookies, resulting in flow features based on exfiltrations and sets being helpful for the classifier.}
  \label{fig:exfilandstorage}
  \vspace{-10pt}
\end{figure}
We conduct feature analysis to understand the most influential features in the classification of cookies. 
We find that the most influential features are the flow features, which capture cookie exfiltrations, set operations, and redirections by cookie setters. 
Figure~\ref{fig:exfilandstorage} shows the distributions for the number of cookie exfiltrations (top) and the number of times a cookie is set (bottom), for \ats and \nonats cookies.
\ats cookies are much more likely to be exfiltrated than \nonats cookies: \ats have a median number of 6 exfiltrations (mean/std is 11.11/15.95) as compared to a median of 0 for \nonats (mean/std is 0.62/5.29).
Also, \ats cookies tend to be set much more frequently by scripts, with a median of 3 set operations (mean/std is 4.86/6.99) as compared to 1 for \nonats cookies (mean/std is 2.17/6.08).

Our analysis of 26,242 first-party ATS cookies which were set more than 3 times on the same site on our crawls shows that 50.2\% percent of these cookies were set with the same value but different expiry values.
This points towards periodically re-setting a cookie being an approach used by trackers to evade expiry limits enforced by ITP (Safari) \cite{Wilander2018ITP2} and ETP (Firefox) \cite{firefox-etp}.
For the rest of the \ats cookies, it appears the most common use case of re-setting is to update the ID value stored in the cookie.
As we described in our threat model and case studies, the ID values stored in these cookies are updated when the \ats obtains more information about the user and finds a new match in the ID graph.
Thus, it is not surprising that these \ats cookies are continuously being updated with new, improved identifiers for the user.

These findings confirm our conclusions in Section~\ref{sec:measurements}: first-party \ats cookies are used to store identifiers which are then exfiltrated to multiple endpoints.

\para{Error analysis.}
We conduct a manual analysis of \name's false positives and false negatives to understand failures. 

We find that the cookies that were most misclassified as \ats are those whose publicly available descriptions indicate they are used to track visitors on a page (e.g., \texttt{\_\_attentive\_id}, \texttt{messagesUtk}, \texttt{omnisendAnonymousID})~\cite{attentive, omnisend, hubspot}.
We also find a few instances of well-known Google Analytics cookies \texttt{\_ga} and \texttt{\_gid} that are labeled in ground truth as \nonats, but are classified by \name as \ats.
Our manual inspection also shows that the false positives are not caused by misclassifications, but mostly that the tracking cookies flagged by \name were mislabeled as \nonats in the ground truth.
In other words, \name has likely correctly classified these tracking cookies. 
We note that even after our procedures to improve ground truth labels, there may be cookies that did not have self-disclosed labels or were served from slightly different scripts (thereby missing our hash-based script matching). 
This is a limitation of our ground-truth, as it relies on either the self-declaration of the cookie purpose or a match between the setting scripts to determine if a cookie is \ats.
We leave the investigation of methods of improving the ground truth labeling to future work.

Regarding false negatives, i.e., \ats cookies missed by \name, we mainly observe two cases.
First, we have the case of finite coverage of encodings.
A representative case is the \texttt{\_pin\_unauth} cookie. 
Its value is double-base64-encoded, which is not included in the list of potential encoding schemes used by \name to detect exfiltration.
These false negatives can be averted by using a more comprehensive list of encoding schemes or by performing full-blown information flow tracking instead of approximating exfiltration flows; however, the latter would come at a performance cost, as we discuss in Section~\ref{subsec:comparison}. 

Second, we have the case of lack of coverage of actions. Our crawl to create the graphs in \name may not capture all possible actions on a webpage. If \name does not capture sufficient activity during webpage execution, some cookies may not be triggered and therefore, the analysis will miss them. 
We further discuss these cases of false negatives in Section \ref{subsec: completeness limitation}

\subsection{Comparison with Existing Countermeasures}
\label{subsec:comparison}
In this section, we compare \name with existing countermeasures that are used to restrict the effect of first-party cookies.

\para{Intelligent Tracking Prevention (ITP)} is used by Safari as a broad countermeasure against online tracking activities.
Under ITP, Safari limits the maximum expiry time of a first-party cookie set through JavaScript and HTTP requests received from IP addresses different from the host website to seven days \cite{safari-default-cookie}. 
In addition, Safari limits this time to only 24 hours for known trackers.\footnote{Firefox also limits the expiry time for cookies set by known trackers to 24 hours \cite{firefox-redirect-tracking}.}
This can be a prudent countermeasure if the first-party tracking cookies were meant to be a storage for the identifier for the repeat visits of the user.
However, as we have shown in the previous section, first-party tracking cookies are shared with a large number of other domains immediately after being set.
This sharing of identifiers among different trackers is meant to enhance their ability to track users across different sites.
Limiting the amount of time that a cookie is set for will not be able to stop this sharing of information, thus proving ineffective in protecting user privacy.

\subsection{Comparison to classifier-based blocking}
Next, we compare \name with state-of-the-art countermeasures against \ats, CookieBlock \cite{DinoUSENIX22CookieBlock} and WebGraph \cite{Siby22WebGraph}, in terms of detection accuracy, website breakage, and robustness.

\para{CookieBlock} \cite{DinoUSENIX22CookieBlock} is a state-of-the-art approach to classify cookies, including advertising/tracking and analytics.
It makes use of both manually curated allow lists and a machine learning classifier, which mainly relies on features based on cookie attributes (cookie names and values).

\para{\webgraph} \cite{Siby22WebGraph} is the state-of-the-art graph-based approach to classify \ats requests.
As WebGraph is not designed to directly classify cookies, we adapt it by identifying \ats resources identified by \webgraph in \cblockcrawl and generating a block list of cookies for each domain set by those resources.
This list is meant to mimic the effect of blocking these resources on \atscookies.

\subsubsection{Detection Accuracy}
Table \ref{tab:performance_classifier_comparison} compares the detection accuracy of \name with CookieBlock and \webgraph.
\name outperforms both approaches in all metrics. 
The superiority in precision indicates that existing countermeasures result in many more false positives than \name.
These additional false positives mean that previous approaches would block functional first-party cookies, potentially affecting user experience.

We also compared \name's performance against popular filter lists \cite{easylist, easyprivacy}.
We found that 52.51\% (834) third-party script domains that set first-party \ats cookies identified by \name are not blocked by filter lists.
Some of the most common examples include \textit{dynamicyield.com}, \textit{pinimg.com}, \textit{auryc.coml}, \textit{tinypass.com}, and \textit{driftt.com}.
Some of the scripts loaded from these domains might be blocked by filter lists, but our tool finds and blocks tracking cookies from scripts that are either missed by filter lists, or are exempted due to breakage issues.
For example, some scripts from \textit{assets.adobedtm.com} are blocked while others scripts are allowed and set \texttt{s\_sq} tracking cookie.
CDNs like CloudFront are a common example of such domains that are used to serve both functional and tracking scripts.
\begin{table}[!t]
	\centering
	\caption{Classification accuracy of \name, \webgraph, and CookieBlock}
	\begin{tabular}[c]{ l c c c}
		\toprule
		\textbf{Classifier} & \textbf{Accuracy} & \textbf{Precision} & \textbf{Recall} \\
		\midrule
		\name                & \accuracy\%              & \precision\%               & \recall\%            \\
		\webgraph            & 79.05\%              & 71.67\%               & 86.17\%            \\
		CookieBlock         & 72.87\%              & 70.73\%               & 80.85\%            \\
		\bottomrule
	\end{tabular}
	\label{tab:performance_classifier_comparison}
	\vspace{-5pt}
\end{table}

\subsubsection{Website Breakage}
We manually analyze the breakage caused by \name, CookieBlock and \webgraph's on 50 sites that are sampled from the 20K sites used in Section \ref{sec:measurements} (25 sites chosen randomly from the top 100 and other 25 from the rest). \footnote{The list of sites used in breakage analysis is available at: https://github.com/cookiegraph/CookieGraph}

We divide our breakage analysis into four categories of typical website usage: navigation (from one page to another), SSO (initiating and maintaining login state), appearance (visual consistency), and miscellaneous functionality (chats, search, shopping cart, etc.).
We label breakage as major or minor for each category: major breakage -- when it is not possible to use the functionality of the site included in any of the aforementioned categories, and minor breakage -- when it is difficult, but not impossible, for the user to make use of the functionality.
To assess breakage, we compare a vanilla Chrome browser (with no countermeasures against first-party cookies) with browsers enhanced with an extension that blocks first-party cookies classified as \ats by \name, enhanced with an extension which blocks all cookies set by resources labeled as \ats by \webgraph, and enhanced with the official CookieBlock extension \cite{cookieblock-github}.
In this analysis, we also include two additional configurations: filter lists \cite{easylist, easyprivacy}, and a Google Chrome with all cookies blocked.
We used two reviewers to perform the breakage analysis to mitigate the impact of biases or subjectivity.
Any disagreements between the reviewers were resolved after careful discussion.

Out of the 50 sites, \name only had major breakage on one site where a cookie popup kept freezing up and preventing navigation around the website due to the deletion of a cookie that stores user preferences.
In contrast, \webgraph, CookieBlock, and filter lists cause major breakage in one of the four categories on at least 6\% of the sites.
For example, \webgraph causes issues with cart functionality on \textit{etsy.com}, complete homepage breakage on \textit{aliexpress.us}, and SSO issues on other sites.
Most of the breakage issues of CookieBlock relate to SSO logins and additional login-dependent functionality (\eg missing profile picture).
Our results, that CookieBlock causes breakage on 10\% of the sites with SSO logins, are similar to the 7-8\% breakage reported by the authors \cite{DinoUSENIX22CookieBlock}.
Blocking all cookies results in major breakage on 32 percent of the sites tested, with SSO and cart functionality proving to be the most recurring issue.

We also find that \webgraph blocks some additional first-party cookies that are important for server-side functionality, but not directly related to user experience and therefore not immediately perceptible. 
For example, \webgraph blocks essential cookies such as \texttt{Bm\_sz} cookie used by Akamai for bot detection, \texttt{XSRF-TOKEN} cookie used to prevent CSRF on different sites, and \texttt{AWSALB} cookies used by Amazon for load balancing.
\name correctly classified these cookies at \nonats, and thus does not prevent these measures from being deployed.
\begin{table}[!htpb]
  \centering
  \caption[]{Website breakage comparison of all three countermeasures.(\greenline) signifies no breakage, (\orangeline) minor breakage, and (\redline) major breakage. Each cell represents the percentage of sites on which breakage was observed.
  }
  \resizebox*{\linewidth}{!}{%
    \begin{tabular}{l  c  c  c  c  c  c  c  c}
      \toprule
      \multirow{2}{*}{\textbf{Classifier}}    & \multicolumn{2}{c}{\textbf{Navigation}} &
      \multicolumn{2}{c}{\textbf{SSO}}        &
      \multicolumn{2}{c}{\textbf{Appearance}} &
      \multicolumn{2}{c}{\textbf{Miscellaneous}}                                                                                                                                                                                                          \\
                                              & Minor                                   & Major                & Minor                & Major                & Minor                & Major                & Minor                 & Major                \\
      \midrule
      \name                                   & \cellcolor{green}0\%                   & \cellcolor{red}2\% & \cellcolor{green}0\% & \cellcolor{green}0\% & \cellcolor{green}0\% & \cellcolor{green}0\% & \cellcolor{green}0\%  & \cellcolor{green}0\% \\
      \webgraph                               & \cellcolor{orange}6\%                   & \cellcolor{red}2\%   & \cellcolor{green}0\% & \cellcolor{red}2\%   & \cellcolor{orange}4\% & \cellcolor{red}2\%   & \cellcolor{orange}2\% & \cellcolor{red}2\%  \\
      CookieBlock                             & \cellcolor{orange}2\%                    & \cellcolor{green}0\% &
      \cellcolor{green}0\%                    & \cellcolor{red}10\%                      & \cellcolor{green}0\% &
      \cellcolor{green}0\%                      & \cellcolor{orange}2\%                    & \cellcolor{red}2\%   \\

      Filter lists                             & \cellcolor{orange}4\%                    & \cellcolor{red}2\% &
      \cellcolor{green}0\%                    & \cellcolor{red}2\%                      & \cellcolor{orange}2\% &
      \cellcolor{red}2\%                      & \cellcolor{orange}2\%                    & \cellcolor{red}4\%   \\ 
      
      No Cookies                             & \cellcolor{orange}8\%                    & \cellcolor{red}8\% &
      \cellcolor{green}0\%                    & \cellcolor{red}32\%                      & \cellcolor{orange}6\% &
      \cellcolor{red}12\%                      & \cellcolor{orange}2\%                    & \cellcolor{red}28\%   \\

      \bottomrule
    \end{tabular}}

  \label{tab:breakge_overview}
  \vspace{-.3in}
\end{table}

\subsubsection{Robustness}
\label{sub:robustness_comparison}
We compare the robustness to evasion of \name, CookieBlock, and \webgraph, \ie to intentional modifications of the cookies to cause the misclassification of \ats cookies as \nonats.
Since \ats are known to engage in the arms race with privacy-enhancing tools \cite{Alrizah19errorsfilterlists,Iqbal17AntiABIMC,hieu2021cv}, it is important to test whether the detection of first-party \ats cookies is brittle in the face of trivial manipulation attempts such as changing cookie names.

We evaluate robustness on a test set of 2,000 sites from our dataset which also have the required CMP needed by CookieBlock for data collection and training.
This translates to a total set of 7,726 first-party cookies.
We change the names of the cookies in our test set to randomly generated strings between 2 and 15 characters.
Both \name and \webgraph are fully robust to manipulation of cookies names while CookieBlock's accuracy degrades by more than 15.87\%, while precision and recall degrade by 15.23\% and 16.79\% respectively. 
\name and \webgraph are robust because they do not use any content features (features related to the cookie characteristics, such as cookie name or domain) since these can be somewhat easily manipulated by an adversary aiming to evade classification \cite{Siby22WebGraph}.
On the contrary, the most important feature of CookieBlock depends on the cookie name, i.e., whether the name belongs to the top 500 most common cookie names \cite{cookieblock-thesis}.

\name's implementation of flow features can be manipulated by an adversary by using a different encoding than it currently considers, or by changing the domains of exfiltration endpoints. 
\name's robustness to these attacks can be improved by more comprehensive information flow tracking. 
However, full-blown information flow tracking would incur prohibitively high run-time overheads (up to 100X-1000X~\cite{Hedin14JSFlowSAC}) and implementation complexity in the browser~\cite{Chudnov2015FlowCCS, Chen18MystiqueCCS, Stock14XSSProtectionUsenix, Lekies13FlowXSSCCS}.
This overhead is significant not only at runtime but also in an offline setting.
Optimistically, assuming a 100X overhead, the time required to crawl a single page increases from 60 seconds to 100 minutes.
Crawling the landing page and 20 internal pages for one website will thus take 34 hours rather than 20 minutes.
In addition to prohibitively large time for website crawls, this delay would also likely impact the fidelity of the page execution itself.

To assess the robustness of \name against manipulation of these flow features, we remove the features related to the flow of cookie information (exfiltration and infiltration of first-party \ats cookies) and then re-train/test the classifier.
We find that \name's accuracy drops by only 2\% when exfiltration and infiltration features are removed.
Our feature analysis using information gain shows that instead of focusing on exfiltration features, \name shifts focus to other features such as the number of local storage accesses by a script and redirections by cookie setters.
While there is a slight performance degradation when these features are removed, \name is able to adapt and still outperforms existing countermeasures by more than 10\% in terms of classification accuracy.

\section{Deployment}
\label{sec:deployment}
We deploy \name to classify first-party cookies in our crawl of 20\% of the top-million sites.

\para{Prevalence of first-party \ats cookies.} 
\name classifies 61.37\% of the 108,947 first-party cookies in our dataset as \ats.
We find that 89.86\% of sites deploy at least one first-party \ats cookie.
Of these sites, the average number of first-party \ats cookies per site is 12.38.

\para{Who sets first-party \ats cookies?} 
The vast majority (96.61\%) of the first-party \ats cookies are set by third-party embedded scripts served from a total of 2,099 unique domains. 
This shows that first-party \ats cookies are in fact set and used by third-parties.
These first-party cookies enable third-parties to perform \textit{\textbf{same-site tracking}} as described in Section \ref{sec:threat_model}.

\para{Who sends and receives first-party \ats cookies?}
Next, we analyze the most prevalent first-party cookies and the third-party entities that actually set them.
Table \ref{tab:top_unlabelled_cookie_domains} lists the top-25 out of 20,794 \atscookies\footnote{We report distinct tuples of the cookie name and the setter script's URL.} based on their prevalence\footnote{Prevalence denotes the percentage out of all sites analyzed on which the cookie was classified as \ats.
Instances where the classification was \nonats are excluded from the prevalence analysis.}.
Two major advertising entities (Google and Facebook) set \atscookies on approximately a third of all sites in our dataset.
\name detects \texttt{\_gid} and \texttt{\_ga} cookies by Google Analytics as \ats on 62.63\% and 53.27\% of the sites.
The public documentation acknowledges using these two first-party cookies to store user identifiers for tracking \cite{google-analytics-cookie-usage}.
We also find evidence of widespread cross-domain first-party \atscookie sharing.
For example, \texttt{\_gid} and \texttt{\_ga} cookies are respectively exfiltrated to 83 and 259 destination domains, more than 95\% of which are non-Google domains. 

\name detects \texttt{\_fbp} cookie by Facebook as \ats on 24.82\% of the sites.
Their public documentation acknowledges that Facebook tracking pixel stores unique identifiers in the first-party \texttt{\_fbp} cookie \cite{fbp-fbc-parameters}.
In fact, Facebook made a recent change to include first-party cookie support in its tracking pixel to avoid third-party cookie countermeasures \cite{facebook-first-party-cookie-cc}.
It is again noteworthy that the \texttt{\_fbp} cookie by Facebook is exfiltrated to 423 destination domains, more than 98\% of which are non-Facebook domains.

TikTok, a social media app that is known to aggressively harvest sensitive user information \cite{tiktok-source-code-report}, also recently added support for setting first-party tracking cookies using TikTok Pixel  \cite{tiktok-first-party-cookie, tiktok-first-party-cookie-adexchanger}.
TikTok's first-party \texttt{\_ttp} tracking cookie is present on 3.69\% percent of sites, which is considerably lower than Facebook and Google but comparable to more specialized entities such as Criteo.
Criteo's \texttt{cto\_bundle} cookie is amongst the most prevalent first-party \ats cookies. 
We observe that Criteo sets this \atscookie on 3.19\% of the sites in our dataset and is exfiltrated to 24 destinations.

The extensive sharing of first-party \ats cookies to other domains enables \textit{\textbf{cross-domain same-site tracking}}, through which a tracker who is unable to set first-party cookies is still able to track user activity on a site.
Similar to results by \cite{Fouad20PixelsPETS}, we find \ats cookies to be extensively shared in redirects. 22\% of all first-party cookies exfiltrated were found to be part of redirects.
Table \ref{tab:top_unlabelled_cookie_domains} highlights extensive cookie-syncing between different \ats, \eg Yandex and Hubspot's first-party \ats cookies are shared to Google Analytics.

\name makes use of these cookie-syncing based characteristics of first-party \ats cookies to detect tracking behavior.
In 15.73\% of cases, the number of exfiltrations by a script setting a first-party \ats cookie is the most important classification feature \footnote{We use treeinterpreter (https://github.com/andosa/treeinterpreter) to determine the most important feature during the classification of \ats cookies.}, while the number of redirects sent is the most important feature in 6.39\% of cases.
Table \ref{tab:top_unlabelled_cookie_domains} lists the most important feature for the classification of each of the top-25 \ats cookies, which shows the importance of both exfiltration and redirect information in detecting first-party \ats cookies.
\begin{table*}[!htpb]
    \centering
    \caption{List of top-25 \ats cookies detected by \name}
   \resizebox{\textwidth}{!}{
   \begin{tabular}{l c c c c c c c c}
   \toprule
        \textbf{Cookie } &          \textbf{Script} & \textbf{Org.} & \textbf{Percentage} & \textbf{Destination} & \textbf{Most Important} & \multicolumn{3}{c}{\textbf{Top-3 Destination Domains}} \\
        \textbf{Name} & \textbf{Domain} & \textbf{} & \textbf{of Sites} & \textbf{Domains} & \textbf{Feature} & \# 1 & \# 2 & \# 3 \\
        \midrule
    \_gid &   google-analytics.com & Google &       62.63 &          83 & LocalStorage Sets through JavaScript  & google-analytics.com  & doubleclick.net & mountain.com \\
    \_ga &   google-analytics.com & Google &      53.27 &         259 & LocalStorage Sets through JavaScript  & google-analytics.com & doubleclick.net & google.com \\
    \_ga &   googletagmanager.com & Google &      31.31 &         222 & LocalStorage Sets through JavaScript & google-analytics.com & doubleclick.net & google.com' \\
    \_fbp &           facebook.net & Facebook &      24.82 &         423 & Exfiltrations through URL & facebook.com & datadoghq.com & google-analytics.com\\
    \_gcl\_au &   googletagmanager.com & Google &      19.05 &          39 & LocalStorage Sets through JavaScript & doubleclick.net & google.com & anytrack.io\\
    \_\_gpi &  googlesyndication.com & Google &      10.06 &           5 & Redirects by Setting Script & doubleclick.net & googleadservices.com & clicktripz.com\\
    \_\_gads &        doubleclick.net &  Google &      9.47 &          11 & Redirects by Setting Script & doubleclick.net & googleadservices.com & wmcdp.io\\
    \_\_gads &  googlesyndication.com & Google &       9.26 &           4 & Redirects by Setting Script & doubleclick.net & googleadservices.com & clicktripz.com\\
    \_\_gpi &        doubleclick.net &  Google &      8.61 &          11 & Redirects by Setting Script & doubleclick.net &  googleadservices.com & wmcdp.io\\
    ln\_or &              licdn.com &  Microsoft &      8.04 &           2 & LocalStorage Sets through JavaScript  & tiqcdn.com & tealiumiq.com & \\
    \_uetsid &               bing.com &  Microsoft &      7.47 &         115 & LocalStorage Sets through JavaScript  & bing.com & clarity.ms & datadoghq.com \\
    \_uetvid &               bing.com &  Microsoft &      7.47 &         134 & LocalStorage Sets through JavaScript  & bing.com & clarity.ms & datadoghq.com \\
    \_ym\_d &              yandex.ru & Yandex &      6.29 &         312 & Redirects by Setting Script & google-analytics.com & doubleclick.net & google.com \\
    \_ym\_uid &              yandex.ru & Yandex &       6.29 &         103 & Redirects by Setting Script & google-analytics.com & adfox.ru & doubleclick.net\\
    \_hjTLDTest &             hotjar.com & HotJar &      6.19 &        1955 & Exfiltrations through URL & google-analytics.com & google.com & facebook.com\\
    \_\_utmz &   google-analytics.com &  Google &     5.12 &           6 & LocalStorage Sets through JavaScript  & google-analytics.com & retargetly.com & zbj.com\\
    \_\_utmb &   google-analytics.com &  Google &      5.12 &          11 & LocalStorage Sets through JavaScript  & google-analytics.com & doubleclick.net & google.com\\
    \_\_utma &   google-analytics.com &  Google &      5.12 &          14 & LocalStorage Sets through JavaScript  & google-analytics.com & thedermreview.com & paltalk.com\\
    \_\_utmc &   google-analytics.com &  Google &      5.01 &          26 & LocalStorage Sets through JavaScript  & google-analytics.com & yandex.ru & moatads.com\\
    OptanonConsent &          cookielaw.org & CookieLaw &       4.04 &           1 & LocalStorage Sets through JavaScript  & gbqofs.io & & \\
    \_clck &             clarity.ms &  Microsoft &      3.97 &           6 & Redirects by Setting Script & ezoic.net & doubleclick.net & tealiumiq.com\\
    \_clsk &             clarity.ms &  Microsoft &      3.93 &           5 & Redirects by Setting Script & smart-bdash.com & tealiumiq.com & brightfunnel.com\\
    \_ttp &             tiktok.com &  TikTok &      3.69 &          19 & LocalStorage Sets through JavaScript  & tiktok.com & tiqcdn.com & uxfeedback.ru\\
    \_\_qca &         quantserve.com & Quantcast &       3.38 &          67 & LocalStorage Sets through JavaScript  & rubiconproject.com & yahoo.com & gumgum.com\\
    cto\_bundle &             criteo.net & Criteo &       3.19 &          24 & Exfiltrations through URL & criteo.com & clarity.ms & akstat.io\\
\bottomrule
\end{tabular}

  }

    \label{tab:top_unlabelled_cookie_domains}
    \vspace{-10pt}
\end{table*}

\para{Cross-site tracking.}
As discussed in Section \ref{sec:threat_model}, trackers use deterministic (\eg email address) or probabilistic (fingerprinting) identifiers for cross-site tracking using first-party cookies.\footnote{While our automated crawls do not allow us to test the use of deterministic identifiers for cross-site tracking at scale, recent work \cite{cartologychang2022ccs} showed the use of email addresses and other deterministic identifiers by trackers such as Criteo.}
We show that scripts that set first-party \ats cookies are also involved in fingerprinting.

First, we analyze the first-party cookies set by the scripts from entities known to engage in browser fingerprinting. 
We use Disconnect's sublist of fingerprinters \cite{disconnect-fingerprinting,firefox-fingerprinting} from its tracking protection list \cite{disconnect-tracking-list}.
We find that 50 (0.42\%) distinct domains that set first-party cookies are also known fingerprinters.
These domains are responsible for setting 32.17\% of all \atscookies.

Second, we use FP-Inspector \cite{Iqbal21FingerprintingSP} to further determine whether first-party \ats cookies are set by fingerprinting scripts.
Using FP-Inspector, we find that fingerprinting scripts set first-party cookies on 1,908 out of 20K sites. 
In total, 632 first-party cookies are set by fingerprinting scripts.
242 out of these 632 cookies, set by 175 different fingerprinting scripts, are classified by \name as \ats cookies. 
It is noteworthy that all of these 242 cookies (e.g., \texttt{adtech\_uid}, \texttt{tfstk}, \texttt{bafp}, \texttt{pxde}, \texttt{ssid}) are not listed as tracking cookies on Cookiepedia. 
Our manual analysis of the remaining 390 \nonats cookies shows that they store non-identifiable information (\eg domain names, flags for cookie permissions).

\section{Limitations}
\label{sec:limitations}
\subsection{Completeness}
\label{subsec: completeness limitation}
\name relies on a graph representation of interactions between different elements during webpage execution.
The number of interactions captured depends on the intensity and variety of user activity on a webpage (\eg scrolling activity, number of internal pages clicked). 
Thus, it is possible that \name does not detect certain \ats cookies if user activity is insufficient as that would mean that its graph representation has not captured particular interactions between different elements in the webpage.

To study the impact of user activity, we recrawl sites performing two to three times more internal page clicks than in the original crawl.
We specifically recrawl 238 sites where Criteo's \texttt{cto\_bundle} cookie was originally classified as \nonats by \name. 
\name's deployment on the recrawled sites results in successful detection of Criteo's \texttt{cto\_bundle} cookie as \ats on 121 of the 238 recrawled sites.
We find that the average number of infiltrations (exfiltrations) increase from 1.54 to 2.95 (1.13 to 4.01) across the original and recrawled sites. We observed a similar trend for other prevalent first-party \ats cookies in our dataset.

We surmise that while there are cases where \name incorrectly classifies \ats as \nonats due to incompleteness of the graph representation, its decision reflects the behavior of the cookie at the time of classification.
As more interaction is captured in the graph, \name is able to correctly switch the label to \ats. More importantly, \name never switches labels from \ats to \nonats due to increased interaction.
\vspace{-10pt}
\subsection{Deployment Overhead}
\name's implementation is not suitable for runtime deployment due to the performance overheads associated with the browser instrumentation and machine learning pipeline.
We envision \name to be used in an offline setting: First first-party \ats cookie-domain pairs are detected using \name and (2) the detected cookie-domain pairs are added to a cookie filter list such as those already supported in privacy-enhancing browser extensions (e.g.,  uBlock Origin \cite{ublock-cookie-remover}) for run-time blocking.
We argue that a reasonably frequent (e.g., once a week) deployment of \name on a large scale would be sufficient in generating and keeping the filter list up-to-date.
This anti-circumvention based approach is frequently used by existing list-based countermeasures and \name's reliance on content features (or lack thereof) prevents evasion by advertisers and trackers. On the other hand, existing countermeasures \cite{DinoUSENIX22CookieBlock}, which heavily make use of cookie name and content features, cannot simply be re-run to generate block lists for updated \ats cookies.
While advertisers and trackers can in theory change cookie names at a rate faster than \name's periodic deployment, updating cookie names frequently is challenging in practice because setting these first-party \ats cookies across many different sites requires tight coordination between different entities. 
To illustrate the practical issues associated with changing cookie names, consider the legacy \texttt{demdex} cookie 
set by Adobe's embedded script that is then exfiltrated to the \textit{demdex.net} domain.
Adobe's documentation explains that it is difficult to change the legacy name because ``... it is entwined deeply with Audience Manager, the Adobe Experience Cloud ID Service, and our installed user base'' \cite{demdex, demdex-change}.
If advertisers or trackers are somehow able to overcome these practical challenges and change cookie names at a much faster pace, \name's online implementation for run-time cookie classification would be necessary. 
Further research is needed for efficient and effective online implementation of \name.
\vspace{-10pt}

\section{Conclusion}
\label{sec:conclusion}
In this paper, we investigated the use of first-cookies for tracking. 
Through a large-scale differential measurement, we showed that trackers use first-party cookies to exfiltrate identifiers even when third-party cookies are blocked. 
We found that third-party cookie blocking is ineffective and blanket first-party cookie blocking is not practical because it results in major functionality breakage on almost one-third of sites.
To detect and block first-party tracking cookies, we proposed \name, a machine-learning approach that captures fundamental tracking behaviors exhibited by first-party cookies. 
Our evaluation showed that \name outperformed the state-of-the-art in terms of detection accuracy, minimization of website breakage, and robustness to evasion attacks.
Our deployment of \name on 20K websites provided evidence of widespread use of first-party tracking cookies on 89.86\% of the tested sites.
These first-party tracking cookies are set by third-party embedded scripts served from 2,099 domains that include major advertising entities such as Google, Facebook, and TikTok.

For reproducibility and to foster follow-up research, \name's source code (patch to OpenWPM and the machine learning pipeline) and the detected list of first-party tracking cookies is available at \textit{https://github.com/cookiegraph/CookieGraph}.


\section*{Acknowledgements}
This work was supported in part by the National Science Foundation under grant numbers 2103439, 2103038, 2138139, and 2127309 (Computing Research Association for the CIFellows 2021 Project). Steven Englehardt was employed by DuckDuckGo during the project, but this work was completed independently.

\bibliographystyle{ACM-Reference-Format}
\bibliography{bibliography}

\appendix
\section{Appendix}

\subsection{Case Studies}
\label{app:case_studies}
In this section, we look at case studies of four popular ATS that make use of first-party tracking cookies: Lotame, ID5, Criteo, and The Trade Desk.

\subsubsection{Lotame}
Lotame is an identity management solution that claims to provide a single ID to users across multiple browsers, devices, and platforms.
Lotame's Lightning Tag~\cite{lotame-lightning-tag} packages the user visit data in a JSON object and sends it to its servers. 
Code \ref{request_lotame} shows an example payload sent to Lotame.
The payload includes IDs assigned by the website, third-party identifiers present on the site, certain user behaviors (configured through collaboration between the publisher and Lotame), and other custom rules defined per website~\cite{lotame-guide}.
Lotame processes the payload and matches the data with its \textit{Cartographer Identity Graph}~\cite{cartographer-identity-graph}, and sends back an ID, called \texttt{panoramaID}~\cite{lotame-panorama-id}, which is stored as a first-party cookie or in localStorage.
\begin{figure}[!b]
\renewcommand{\lstlistingname}{Code}
\begin{lstlisting}[style=htmlcssjs, captionpos=b, label={request_lotame}, caption={Example of data sent structure sent to Lotame during a user's first visit.},captionpos=b, numbers=left, xleftmargin=0em]
data: {
  behaviorIds: [1,2,3],
  behaviors: {
    int: ['behaviorName', 'behaviorName2'],
    act: ['behaviorName']
  },
  ruleBuilder: {
    key1: ['value 1a', 'value 1b']
  },
  thirdParty: {
    namespace: 'NAMESPACE',
    value: 'TPID_VALUE'
  }
}
\end{lstlisting}
\vspace{-10pt}
\end{figure}
\subsubsection{ID5 Universal ID}
\begin{figure}[!t]
\renewcommand{\lstlistingname}{Code}
\begin{lstlisting}[style=htmlcssjs, captionpos=b, label={request_id5}, caption={Example of data structure received from ID5 during a user's first visit.},captionpos=b, numbers=left, xleftmargin=1.5em]
{
   "created_at":"2022-02-09T11:42:40.817811Z",
   "id5_consent":true,
   "original_uid":"ID5*FnFOGLkYzdJ...Oeg2Ok4VTNc",
   "universal_uid":"ID5*HGH7W7iMpMu3-...szRCJDUkiiu-tv5BQ",
   "signature":"ID5_Ab6tnGgm...JQWlsUEfynB1hBGZc",
   "link_type":1,
   "cascade_needed":true,
   "privacy":{
      "jurisdiction":"other",
      "id5_consent":true
   }
}
\end{lstlisting}
\vspace{-10pt}
\end{figure}
ID5 provides identity resolution for publishers and advertisers through its \textit{Identity Cloud}~\cite{id5-identity-cloud}.
ID5's script sends a request to its Identity Cloud with a payload that contains several deterministic identifiers, such as email, usernames, and phone numbers (if available) as well as probabilistic identifiers, such as IP address, user agent, and location of the user \cite{id5}.
Identity Cloud processes and returns an ID, called \textit{universal\_id}, which is stored as a first-party cookie as well as in local storage. 
An example payload from ID5 is shown in Code \ref{request_id5}.
We note that ID5 also provides \textit{Partner Graph}, a service that enables information sharing among its partners \cite{id5-identity-cloud}. 
Partner Graph allows different identity providers to exchange information with each other. 

\subsubsection{Criteo}
Criteo provides \textit{Criteo Identity Graph} for identity resolution \cite{criteo-online-identification}.
Criteo Identity Graph is built from four different sources: (i) data contributed by advertisers, (ii) data collected from publisher websites, (iii) data provided by Criteo partners such as LiveRamp and Oracle, (iv) and predictions on existing data by Criteo's machine learning models.
Criteo claims that its identity graph is able to stitch together identifiers from more than 2 billion users across the world and that it contains persistent deterministic identifiers for 96\% of the users~\cite{criteo-online-identification}.
Similar to other identity resolution services, Criteo generates an ID, based on identifiers, such as hashed emails, mobile device IDs, and cookie IDs, and stores it in both first-party cookies and localStorage as \texttt{cto\_bundle}.
As described in Section \ref{subsec:graph}, \name's graph representation abstracts storage to refer to both Cookies and localStorage, and it includes a count of localStorage accesses in the feature set computed from the graph representation to effectively model this particular behavior.

\subsubsection{The Trade Desk} The Trade Desk (TTD) is a digital marketing company whose stated aim is to improve digital advertising.
Their most relevant initiative is Unified ID 2.0 (UID 2.0) \cite{unifiedid}, which uses deterministic information such as email address and probabilistic information such as browser/device attributes to create identifiers at the household and individual level.
UID 2.0 is unique because of its partnerships with major players and publishers in the digital advertising ecosystem.
Notably, ID5 \cite{ttd-id5} and LiveRamp \cite{ttd-liveramp}, which specialize in providing alternatives to third-party cookie-based tracking, both collaborate with TTD to integrate with UID 2.0.
UID 2.0 works by first collecting hashed email addresses and other deterministic identifiers from users visiting a website, which is then sent to a UID 2.0 operator.
The operator matches the hashed email address with the centralized ID graph consisting of information contributed by all UID 2.0 partners.
In case of a match, an encrypted user identifier (or token) is sent back to the client-side and stored in a first-party cookie.
This token is used by TTD's partners, alongside other deterministic and probabilistic signals, to identify a user through identity graphs as described in section \ref{sec:threat_model}.

\end{document}